\documentclass[smallextended]{article}

\usepackage{amssymb}
\usepackage{amsmath}
\usepackage{amsthm}

\usepackage{bm}

\usepackage{graphicx}
\usepackage[a4paper,margin=2.5cm]{geometry}

\usepackage{enumerate}
\usepackage{bussproofs}

\usepackage{parskip}

\usepackage{etoolbox}

\patchcmd{\thebibliography}{\section*{\refname}}{}{}{}

\title{Field-free electrodynamics}
\date{}
\author{Mischa Moerkamp \\ \footnotesize{\textbf{e-mail:} mdmoerkamp@gmail.com}}

\begin{document}

\maketitle

\begin{abstract}

\noindent
The Maxwell-Lorentz theory of electrodynamics cannot readily be applied to a system of point charges: the electromagnetic field is not well-defined at the position of a point charge, an energy conservation argument is not obvious, an infinite regression arises when the interactions occur along the light cones and the advanced potentials lead to an apparent breakdown of causality. A rather controversial solution to these problems involves instantaneous action at a distance, which comes at the expense of violating Lorentz covariance. Some experiments call into question the applicability of the standard retardation constraint to all components of the electromagnetic field. In light of these experimental results, this paper develops two instantaneous action at a distance theories of electrodynamics, which are compatible with some basic features of classical electrodynamics: the Lorentz transformed Coulomb's law, the Biot-Savart force law and Faraday's law of induction.\\

\noindent
\textbf{Keywords:} Instantaneous action at a distance; Tangherlini transformation; field-free interaction Lagrangian; bound electric field; Weber-type force.
\end{abstract}

\newpage
\tableofcontents

\newpage
\section{Introduction}

\noindent
The Maxwell-Lorentz theory of electrodynamics consists of two separate parts:
\begin{enumerate}
\item Maxwell's equations describe the time evolution of the electric field $  \textbf{E}  $ and the magnetic field $   \textbf{B}  $ generated by an electric charge density $ \rho_e $ and an electric current density $ \textbf{J}_e  $:
$$  \nabla \cdot  \textbf{E}    =  \frac{\rho_e }{ \varepsilon_0 } \hskip \textwidth minus \textwidth  \text{(Gauss's law)}  $$
$$   \nabla \cdot  \textbf{B} = 0   \hskip \textwidth minus \textwidth   \text{(Gauss's law for magnetism)}  $$
$$    \nabla \times  \textbf{B}  -  \frac{1}{c^2}  \frac{ \partial   \textbf{E}}{\partial t}  =     \mu_0   \textbf{J}_e    \hskip \textwidth minus \textwidth    \text{(Amp\`{e}re's circuital law)}   $$
$$    \nabla \times  \textbf{E}  + \frac{ \partial    \textbf{B}     }{\partial t} = 0    \hskip \textwidth minus \textwidth   \text{(Maxwell-Faraday equation)}  $$

\noindent
Initial and boundary conditions are required to uniquely specify the electromagnetic (EM) field.

\item The Lorentz force law describes the EM force acting on a particle with an electric charge $q_e$ and a velocity $  \textbf{u} $:
$$    \textbf{F}   =     q_e  (    \textbf{E}   +    \textbf{u}    \times   \textbf{B}    )          \hskip \textwidth minus \textwidth  \text{(Lorentz force law)}   $$
Recall that the force $\textbf{F}$ is defined as the time derivative of the linear momentum:
$$     \textbf{F}  =    \frac{d  }{d t} (   \gamma [ \textbf{u} ]   m_0  \textbf{u}   )   =  m_0   \gamma [    \textbf{u} ]    \left(   \textbf{a}   +          \gamma [  \textbf{u}  ]^2      \boldsymbol{\beta}     \left(  \boldsymbol{\beta}   \cdot     \textbf{a}   \right)        \right)  \hskip \textwidth minus \textwidth   \text{ }   $$
where $ \gamma [  \textbf{u}  ] = 1/ \sqrt{ 1 -  |  \textbf{u}  |^2 / c^2  } $ is the Lorentz factor, $  m_0    $ is the rest mass, $  \textbf{a} $ is the acceleration and $ \boldsymbol{\beta} = \textbf{u} /c $ is the proper velocity. If one inverts this to compute the acceleration, one obtains:
$$   \textbf{a} = \frac{   \textbf{F} -   \boldsymbol{\beta}     (   \boldsymbol{\beta}   \cdot    \textbf{F}   )     }{m_0 \gamma[   \textbf{u}  ]}      \hskip \textwidth minus \textwidth   \text{ }   $$
\end{enumerate}

\noindent
From Maxwell’s equations, it can be shown that disturbances of the EM field propagate at the speed of light. The EM field can be expressed in terms of the scalar potential $\phi$ and the vector potential $  \textbf{A} $:

\begin{minipage}[t]{0.3\textwidth}
$ \textbf{E} = - \nabla \phi - \dfrac{\partial   \textbf{A}  }{\partial t }  $
\end{minipage}
\begin{minipage}[t]{0.7\textwidth}
$  \textbf{B}  = \nabla \times   \textbf{A}     \hskip \textwidth minus \textwidth   \text{ }  $
\end{minipage}

\noindent
and Maxwell's equations can be written in potential form:
$$ \nabla^2  \phi + \frac{\partial }{\partial t} (\nabla \cdot     \textbf{A}   ) = -  \frac{  \rho_e }{  \varepsilon_0 }    \hskip \textwidth minus \textwidth   \text{ } $$
$$      \left(  \nabla^2   \textbf{A} - \frac{1}{c^2 } \frac{\partial^2   \textbf{A}   }{\partial t^2 }  \right) - \nabla \left( \nabla \cdot   \textbf{A} + \frac{1}{c^2 } \frac{\partial \phi}{\partial t}   \right)  =  - \mu_0      \textbf{J}_e  \hskip \textwidth minus \textwidth   \text{ } $$
For any choice of a twice-differentiable scalar function $ f $ of position and time, the following transformation can be made without changing the EM field:

\begin{minipage}[t]{0.3\textwidth}
$  \phi  \to     \phi - \dfrac{\partial f }{\partial t}   $
\end{minipage}
\begin{minipage}[t]{0.7\textwidth}
$    \textbf{A}  \to      \textbf{A} + \nabla  f     \hskip \textwidth minus \textwidth   \text{(Gauge freedom)}  $
\end{minipage}

\noindent
One such gauge is the Lorenz gauge, in which case $f$ is chosen such that:
$$   \nabla \cdot   \textbf{A}  +  \frac{1}{c^2}  \frac{\partial \phi  }{\partial t}  = 0  \hskip \textwidth minus \textwidth  \text{(Lorenz gauge condition)}  $$
The Lorenz gauge results in the following Maxwell equations in potential form:
$$  \frac{1}{c^2} \frac{\partial^2  \phi }{\partial t^2 } -  \nabla^2  \phi  =  \frac{ \rho_e }{  \varepsilon_0 }   \hskip \textwidth minus \textwidth   \text{ }  $$
$$    \frac{1}{c^2 }  \frac{ \partial^2   \textbf{A}  }{\partial  t^2 } - \nabla^2    \textbf{A}  =  \mu_0   \textbf{J}_e    \hskip \textwidth minus \textwidth   \text{ } $$
These equations yield the retarded and advanced solutions (corresponding to $+$ and $-$ respectively):
$$ \phi_{\pm} [    \textbf{x} , t ]   = k_e  \int  \frac{  \rho_e \left[   \textbf{x}' , t'  \right]  }{|   \textbf{x} -   \textbf{x}' |}  \delta \left[ t'   - t_\pm \right]  d    \textbf{x}'  d t'   \hskip \textwidth minus \textwidth   \text{ }  $$
$$      \textbf{A}_{\pm} [   \textbf{x} ,t ]   = k_m  \int  \frac{     \textbf{J}_e [    \textbf{x}' ,t'  ]  }{|  \textbf{x} -   \textbf{x}' |}   \delta \left[ t'  - t_\pm  \right]   d    \textbf{x}' dt'   \hskip \textwidth minus \textwidth   \text{ }  $$
where the retarded and advanced times are: $ t_{\pm} = t  \mp    |   \textbf{x} -   \textbf{x}' | / c$. A direct application of the Maxwell-Lorentz theory to a system of point charges is hindered by a number of problems. A possible solution to these problems involves instantaneous action at a distance (IAAD), which comes at the expense of violating Lorentz covariance. IAAD cannot be excluded based on conventional optical experiments, because there is an alternative constructive interpretation of these experiments in terms of IAAD forces that are dependent on the velocity of the measuring device relative to a preferred reference frame (PRF). Furthermore, we point out some empirically based arguments for reconsidering IAAD as a genuine possibility. Lastly, this paper develops two instantaneous action at a distance theories (IAADTs) of electrodynamics, which are compatible with some basic features of classical electrodynamics: the Lorentz transformed Coulomb's law, the Biot-Savart force law and Faraday's law of induction.

\newpage
\section{Maxwell-Lorentz theory in a system of point charges}

\noindent
The shortcomings of the Maxwell-Lorentz theory in a system of point charges include:

\begin{enumerate}
\item The limit of the EM field at the position of a point charge does not exist.
\item The energy stored inside the EM field generated by a point charge diverges, so an energy conservation argument is not obvious in a system of point charges. Furthermore, accounting for the radiation recoil force yields runaway and acausal behaviour.
\item A closed set of equations cannot be obtained if the world lines are described by means of interactions along the light cones.
\item The advanced potentials lead to an apparent breakdown of causality.
\end{enumerate}

\noindent
\textbf{Problem 1:} A precise definition of the EM field at the position of a point charge is hindered by the divergence of the near EM field. Let $  \textbf{r} $ be the position of a point charge $q_e \neq 0$ and let $B [  \textbf{r} , \epsilon ]$ be a ball of radius $\epsilon >0$ centered at $  \textbf{r} $. From Gauss's law and the divergence theorem, it follows that:
$$ \lim_{\epsilon \to 0}  \left|  \int_{\partial B [   \textbf{r} , \epsilon ] } \frac{   \textbf{E} \cdot   d   \textbf{A}  }{4 \pi \epsilon^2}    \right| = \lim_{\epsilon \to 0} \left|  k_e \frac{q_e}{  \epsilon^2  } \right| = \infty \hskip \textwidth minus \textwidth   \text{ }   $$
Hence, the average length of the radial component of $ \textbf{E}$ relative to the point $ \textbf{r}$ diverges if $\epsilon$ goes to zero. However, we can study the behaviour of the EM field outside of a point charge.

\noindent
\textbf{Problem 2:} The EM field can hold energy, linear and angular momentum of its own, on par with the mechanical masses. The EM energy density $ u_{em} $ and the EM Poynting vector $   \textbf{S}_{em}$ are given by:
$$  u_{em}  =   \dfrac{1}{2} \left( \varepsilon_0    |  \textbf{E} |^2  +   \dfrac{1}{ \mu_0 }  |  \textbf{B} |^2   \right)   \hskip \textwidth minus \textwidth   \text{ }  $$
$$    \textbf{S}_{em} =  \dfrac{1}{ \mu_0}   \textbf{E}   \times    \textbf{B}  \hskip \textwidth minus \textwidth   \text{ }  $$

\noindent
and the EM version of Poynting's theorem is:
$$ \frac{ \partial    u_{em}}{ \partial t}  +    \nabla \cdot   \textbf{S}_{em}    +   \textbf{J}_e  \cdot   \textbf{E} = 0  \hskip \textwidth minus \textwidth   \text{ }   $$
However, the energy stored inside the EM field generated by a stationary point charge diverges:
$$ E_{em} = \frac{  \varepsilon_0}{2}    \int    |   \textbf{E} |^2    d   \textbf{x}   =   k_e \int   \frac{ q_e^2  }{ 8  \pi   r^4 }  d  \textbf{x}  = \infty  \hskip \textwidth minus \textwidth   \text{ }    $$
so an energy conservation argument is not obvious in a system of point charges. One may use the retarded potentials:
$$ \phi_+  [    \textbf{x}  , t  ]    =   k_e         \int  \frac{   q_e  }{   |   \textbf{x}   -   \textbf{r}  [t'] |   }   \delta \left[     t'   +   \frac{    |   \textbf{x}   -   \textbf{r}   [t'] |       }{c}   -   t     \right]    d  t '      \hskip \textwidth minus \textwidth   \text{ }  $$
$$     \textbf{A}_+  [   \textbf{x} , t ]  =  k_m      \int \frac{ q_e      \textbf{u}   \left[  t'  \right]    }{  |   \textbf{x} -   \textbf{r}   [t'] |   }       \delta   \left[  t' +  \frac{     |   \textbf{x} -   \textbf{r}   [t'] |      }{c} - t  \right]    d t'   \hskip \textwidth minus \textwidth   \text{ }  $$
to obtain the retarded EM field of a point charge:
$$     \textbf{E}_+  [   \textbf{x}  , t  ] =         \underbrace{   k_e   q_e   \frac{   (  \hat{\textbf{n}}_+   -  \boldsymbol{\beta}_+   ) (  1 -   | \boldsymbol{\beta}_+   |^2    )  }{    ( 1 -   \hat{\textbf{n}}_+   \cdot \boldsymbol{\beta}_+   )^3 |    \textbf{x}  -   \textbf{r}_+  |^2  }  }_\text{Velocity-dependent electric field}  +     \underbrace{  k_e     q_e    \frac{   \hat{\textbf{n}}_+   \times  ( (   \hat{\textbf{n}}_+   - \boldsymbol{\beta}_+   )   \times \dot{\boldsymbol{\beta}}_+    )  }{c   ( 1 -  \hat{\textbf{n}}_+   \cdot \boldsymbol{\beta}_+  )^3  |  \textbf{x}  -   \textbf{r}_+ |  }   }_\text{Acceleration-dependent electric field}      \hskip \textwidth minus \textwidth   \text{ }  $$
$$   \textbf{B}_+   [  \textbf{x}    , t  ]  =                \underbrace{ k_e  q_e   \frac{   \hat{\textbf{n}}_+   \times    (  \hat{\textbf{n}}_+   -  \boldsymbol{\beta}_+  ) (  1 -   | \boldsymbol{\beta}_+  |^2    )  }{ c    ( 1 -  \hat{\textbf{n}}_+  \cdot \boldsymbol{\beta}_+  )^3 |    \textbf{x}  -   \textbf{r}_+   |^2  }   }_{\text{Velocity-dependent magnetic field}}  +  \underbrace{   k_e     q_e   \frac{  \hat{\textbf{n}}_+   \times  (   \hat{\textbf{n}}_+    \times  ( (  \hat{\textbf{n}}_+   - \boldsymbol{\beta}_+  ) \times \dot{\boldsymbol{\beta}}_+   )  )  }{ c^2    ( 1 -  \hat{\textbf{n}}_+   \cdot \boldsymbol{\beta}_+  )^3  |  \textbf{x}  -   \textbf{r}_+    |  }    }_{\text{Acceleration-dependent magnetic field}}        =   \frac{      \hat{\textbf{n}}_+     \times  \textbf{E}_{   + }    [  \textbf{x}  ,   t  ] }{c}     \hskip \textwidth minus \textwidth   \text{ }  $$
where $      \hat{\textbf{n}}_+    = (  \textbf{x} -    \textbf{r}_+  ) /  |   \textbf{x}   -  \textbf{r}_+  |   $, $  \boldsymbol{\beta}_+     =   \textbf{u} [ t_+  ] / c    $ and $ \textbf{r}_+ =   \textbf{r} [ t_+ ] $. It can subsequently be shown that the radiated EM power of a point charge equals:\cite{jackson}
$$  P =  k_e  \frac{2}{3}  \frac{ q_e^2  \gamma [   \textbf{u}   ]^6   }{ c^3 }  \left(  |  \textbf{a} |^2  - |   \boldsymbol{\beta}  \times   \textbf{a}  |^2   \right)   \hskip \textwidth minus \textwidth   \text{(Generalized Larmor formula)}    $$
Once we take the resulting radiation recoil force into account, the point charge exhibits runaway and acausal behaviour. Suppose that $  |  \textbf{u}  |  \ll c$, then the energy loss due to the self-force $  \textbf{F}_\text{recoil}$ obeys the following equation:
\begin{flalign}\notag
    \int_{t_1}^{t_2}   \textbf{F}_\text{recoil}  \cdot   \textbf{u}  dt  & =    - k_e  \frac{2}{3}  \frac{ q_e^2   }{ c^3 }  \int_{t_1}^{t_2} 
   \frac{d  \textbf{u}  }{dt} \cdot  \frac{ d   \textbf{u} }{ dt }  dt       = &&  \\\nonumber
 & =    k_e  \frac{2}{3}  \frac{ q_e^2   }{ c^3 }   \int_{t_1}^{t_2}    \frac{ d   \textbf{a} }{d t }   \cdot   \textbf{u}    dt   &&
\end{flalign}
We can identify the self-force:
$$     \textbf{F}_\text{recoil}  =  k_e  \frac{2}{3}  \frac{ q_e^2   }{ c^3 }   \frac{ d   \textbf{a} }{dt}      \hskip \textwidth minus \textwidth   \text{ }    $$
and the complete equation of motion for the point charge is given by:
$$    m    \textbf{a} =   q_e  ( \textbf{E}  +    \textbf{u}   \times  \textbf{B}     )  +    \textbf{F}_\text{recoil}         \hskip \textwidth minus \textwidth   \text{ }    $$
Consider a particle in one dimension with an external force $F [t]$ applied to it, then:

\begin{minipage}[t]{0.3\textwidth}
$      a -    \tau \dot{a}   =  \dfrac{  F [t]}{m}    $
\end{minipage}
\begin{minipage}[t]{0.7\textwidth}
$      \tau = \dfrac{2}{3}   \dfrac{ k_e  }{ c^3 }  \dfrac{    q_e^2  }{m}     \hskip \textwidth minus \textwidth   \text{ }  $
\end{minipage}

\noindent
which implies that the acceleration takes the form:
$$  a[t]  =  \underbrace{ a_0   \exp \left[ \frac{t}{  \tau } \right]  }_\text{Runaway solution}   +  \underbrace{  \frac{1}{ \tau }  \int_t^\infty   \exp \left[ \frac{ t - t' }{ \tau }  \right]  \frac{F [ t']}{m}  dt'  }_\text{Acausal solution}   \hskip \textwidth minus \textwidth   \text{ }    $$
Choosing $a_0 = 0 $ eliminates the runaway solution. However, the acausal solution cannot be discarded, which implies that the acceleration $ a [t]$ depends on the force $F[t']$ at any later time $t ' > t$. We mention some attempts that have been made to deal with the problem of energy conservation (although this list does not claim to be exhaustive):

\begin{enumerate}[-]
\item Some physicists - most notably Dirac - have tried to regularize the infinite expression.\cite{dirac} The infinite EM energy is compensated by a negatively infinite mechanical mass, which renders the total mass finite.
\item Instead of assuming that the EM field is retarded, Feynman and Wheeler proposed a theory in which the EM field is half-retarded plus half-advanced, so that the particles are not self-interacting.\cite{feynman1,feynman2}
\item It is possible to consider a body whose electric charge and total mass go to zero in an asymptotically self-similar manner.\cite{wald}
\item Other physicists have attempted to fundamentally modify the laws of classical electrodynamics. Two such modifications are the Born-Infeld theory and the Bopp-Podolsky theory, both of which introduce new hypothetical scale parameters.\cite{born-infeld,bopp}
\end{enumerate}

\noindent
\textbf{Problem 3:} Suppose that the EM field at the position $ \textbf{r}_j $ of particle $j$ is given by:
$$   \textbf{E}  [   \textbf{r}_j  , t  ] =     \sum_{ k \neq j }    \Big(   a_+   \textbf{E}_{k + }  [  \textbf{r}_j  ,  t ]  +  a_-    \textbf{E}_{ k - }  [ \textbf{r}_j , t ]   \Big)           \hskip \textwidth minus \textwidth   \text{ }  $$
$$   \textbf{B}  [   \textbf{r}_j  , t  ] =     \sum_{ k \neq j }     \Big(   a_+      \textbf{B}_{k+}  [ \textbf{r}_j ,  t ]      + 
 a_-     \textbf{B}_{k-}   [  \textbf{r}_j  ,  t ]     \Big)           \hskip \textwidth minus \textwidth   \text{ }  $$
where $a_+$ and $a_-$ are constants for which $  a_+ +  a_-  = 1  $. The EM fields $ ( \textbf{E}_{k+} , \textbf{B}_{k+} ) $ and $ ( \textbf{E}_{ k - }  , \textbf{B}_{ k - } ) $ are the retarded and advanced EM fields emanating from particle $k$, respectively.\\

\begin{figure}[h]
\centering
\includegraphics[scale=0.7]{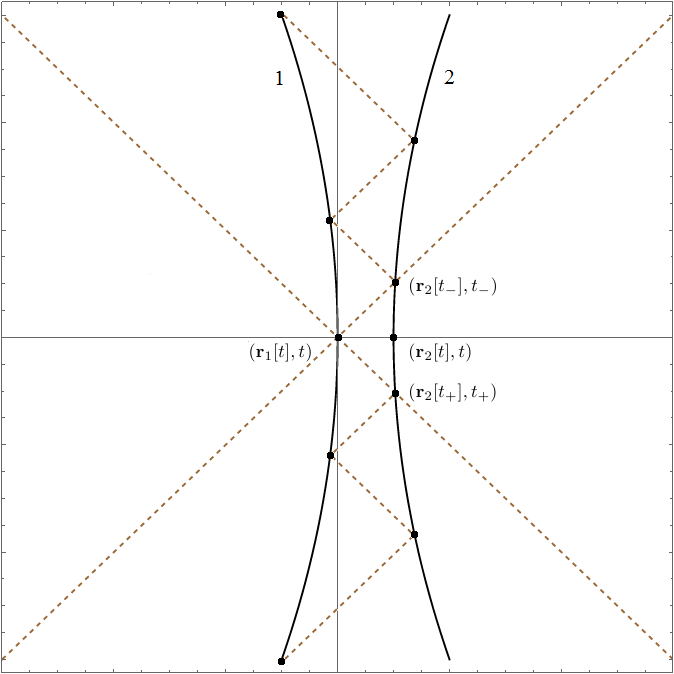}
\caption{The two curves represent the worldlines of the two particles. The dashed lines represent signals that propagate at the speed of light. The light cones emanating from $  (  \textbf{r}_1 [   t  ] ,   t     )  $ intersect the worldline of particle $2$ at $ (  \textbf{r}_2 [ t_+  ] , t_+   ) $ and $(  \textbf{r}_2 [ t_-   ] , t_-   )$.}
\label{lightcones}
\end{figure}

\noindent
Consider particles $1$ and $2$, whose worldlines are shown in Figure \ref{lightcones}. The EM field at $ (   \textbf{r}_1 [ t ]  ,  t ) $ is determined by the dynamical variables of particle $2$ at $ (   \textbf{r}_2 [ t_+  ] , t_+    )$ or $ (   \textbf{r}_2  [ t_- ] , t_-   )  $ (or both). One is caught in an infinite regression, because the EM field at $(  \textbf{r}_2 [ t_+  ] , t_+   ) $ is also determined by its past light cone. Similarly, the EM field at $ (  \textbf{r}_2 [ t_-   ] , t_-    )  $ is determined by its future light cone. Hence, if the world lines are described by means of interactions along the light cones, a closed set of equations cannot readily be obtained. The tentative conclusion can be drawn that the Maxwell-Lorentz theory is not posed as an initial value problem in which the initial conditions are the instantaneous positions and velocities of the particles (the Cauchy data), but rather as a theory in which the initial conditions comprise entire segments of trajectories. A formal proof of a no-interaction theorem for two particles in $3+1$ dimensions was obtained by Currie, Jordan and Sudarshan in 1963, which states that a nondegenerate Hamiltonian theory cannot describe any interaction if the theory is Lorentz covariant.\cite{currie1,currie2} Cannon and Jordan extended this no-interaction theorem to three particles,\cite{cannon} and Leutwyler generalized it to an arbitrary number of particles.\cite{leutwyler}

\noindent
\textbf{Problem 4:} The advanced potentials are difficult to reconcile with the principle of causality. To illustrate this problem, consider the particles shown in Figure \ref{lightcones}. The retarded EM field corresponds to signal propagation forward in time, whereas the advanced EM field represents signal propagation backward in time. Since $  (  \textbf{r}_2 [ t_-   ] , t_-    ) $ lies in the future of $  (  \textbf{r}_1 [ t  ] ,   t    ) $ regardless of the choice of the inertial reference frame (IRF), one concludes that if there is a contribution of the advanced field ($a_-   \neq 0$), then the future affects the present leading to an apparent breakdown of causality. Hence, the advanced Green's function is commonly rejected, which is also referred to as the Sommerfeld condition. The Wheeler-Feynman absorber theory, on the other hand, rejects the Sommerfeld condition and assumes a half-retarded plus half-advanced EM field:\cite{feynman1,feynman2}
$$      \textbf{E}_{\text{tot}}  [  \textbf{r}_j , t]  =    \sum_{ k \neq j }  \frac{  \textbf{E}_{ k + } [  \textbf{r}_j , t ]  +   \textbf{E}_{ k -} [  \textbf{r}_j , t ]   }{2}      \hskip \textwidth minus \textwidth   \text{ }  $$
If the source-free field is assumed to vanish:
$$   \textbf{E}_{\text{free}}  [  \textbf{r}_j , t ]  =   \sum_{ k }  \frac{   \textbf{E}_{ k +}  [   \textbf{r}_j , t ] -   \textbf{E}_{ k - } [    \textbf{r}_j , t  ]    }{2}  = 0          \hskip \textwidth minus \textwidth   \text{(Ansatz Wheeler-Feynman theory)}  $$
then the radiation recoil field may be obtained without the need for self-interaction:
$$      \textbf{E}_{\text{tot}} [   \textbf{r}_j , t]    =   \frac{   \textbf{E}_{j+}  [  \textbf{r}_j , t]   -   \textbf{E}_{ j - }  [   \textbf{r}_j , t ]    }{2} +   \sum_{ k \neq j }   \textbf{E}_{k +} [   \textbf{r}_j , t]       =    \textbf{E}_\text{recoil}  [  \textbf{r}_j , t ]  +   \sum_{ k \neq j }    \textbf{E}_{k +} [  \textbf{r}_j , t]        \hskip \textwidth minus \textwidth   \text{ }  $$
As pointed out before, the radiation recoil field comes with its own set of challenges.

\newpage
\section{Instantaneous action at a distance}

\subsection{Instantaneous action at a distance forces and radiative processes}

\noindent
The problems associated with a system of point charges can be overcome with IAAD:
\begin{enumerate}
\item The IAAD forces between point particles are well-defined.
\item There is no diverging energy problem since we are dealing with instantaneous interparticle interactions. The radiative processes would have to be dealt with separately.
\item If IAAD is employed, then there is no infinite regression. Consider the two particles in Figure \ref{lightcones}. The force acting on particle $1$ at $ (  \textbf{r}_1  [ t ]   ,    t  )  $ is written in terms of the dynamical variables of particle $1$ and particle $2$ at $ (  \textbf{r}_1  [  t  ]   ,   t   ) $ and $ (  \textbf{r}_2  [  t ]   ,   t  ) $ respectively; the force acting on particle $2$ at $ (  \textbf{r}_2  [  t  ]   ,   t    ) $ is also written in terms of these variables. Hence, IAAD yields a closed set of equations and presents an ordinary initial value problem in which the initial conditions are the Cauchy data.
\item IAAD does not pose an immediate threat to causality. It is sometimes claimed that IAAD does violate causality, due to the relativity of simultaneity implied by the Lorentz transformation (LT). An appeal to the LT, however, begs the question, because IAAD comes at the expense of violating Lorentz covariance.
\end{enumerate}

\noindent
In an IAADT, the IAAD forces are described by a field-free Lagrangian, which depends on the instantaneous dynamical variables of the interacting particles. The radiative processes (which involve the real photons) have to be dealt with separately and are constrained by the speed limit. This can be summarized in the following table:

\noindent
\begin{tabular}{ | p{4cm} | p{5cm} | p{5cm} | }
    \hline
     &   Propagation speed in vacuo: &  Carrier:  \\ \hline
  IAAD forces:   &   $ \infty $  &   N/A \\ \hline
    Radiative processes: &   $   c $ &    Photons    \\ \hline
    \end{tabular}

\subsection{Axiomatic foundations of special relativity theory}

\noindent
Before we proceed further, let us make a quick note about the special status of the relativity principle in the axiomatic foundations of special relativity theory (SRT). The standard postulates of SRT are the relativity principle and the light speed invariance (although some additional assumptions are left implicit).\cite{einstein} Ignatowsky, Frank and Rothe showed that a transformation equivalent to the LT, up to a free parameter $ c' $ (which represents the invariant speed), can be obtained from the relativity principle, without making any reference to the light speed invariance.\cite{ignatowsky1}-\cite{frank-rothe} It is known experimentally that the invariant speed $c'$ of the resulting transformation must be extremely close to the speed of light.

\subsection{Constructive and principle theories}

\noindent
In a notable letter to the London Times entitled \textit{"What is the Theory of Relativity?"} published in 1919, Einstein distinguished principle theories from constructive theories:

\begin{enumerate}
\item A constructive theory attempts to construct the more general phenomena by starting out from a simple formal scheme.

\item On the other hand, a principle theory is not constructed, but empirically discovered. 
\end{enumerate}

\noindent
SRT gives an elegant framework that can explain a wide range of experimental results, but it is a metatheory to which the physical laws must conform. Hence, SRT belongs to the class of principle theories and the constructive counterpart of SRT is the underlying Lorentz covariant mechanics. There is an alternative constructive interpretation of conventional optical experiments in terms of forces that are dependent on the velocity of the measuring device relative to a PRF. This interpretation - which we refer to as the Lorentzian interpretation - relaxes the condition of Lorentz covariance. In SRT, our apparent inability to detect any absolute motion is elevated to the status of a postulate, but Lorentz argued that the relativity principle should not be viewed as a postulate, it should rather be viewed as a hypothesis, framed on an experimental basis, and always open to refutation.\cite{lorentz} The Lorentzian interpretation is equivalent to SRT if the assumption of Lorentz covariance is correct and many physicists have therefore come to the conclusion that only SRT should be retained, because it makes the fewest assumptions (as per Ockham's razor). It goes without saying that the simplest or most elegant theory is not guaranteed to be the correct one. Furthermore, the Lorentzian interpretation is a more cautious take on the experimental evidence, because it leaves open the possibility that some forms of information are transmitted faster than the speed of light: any experiment that indicates a faster than light signal invalidates SRT (unless causality is violated), but leaves the Lorentzian interpretation intact. And although Ockham's razor is often invoked to discard the Lorentzian interpretation because of the advertised simplicity of SRT, the Lorentzian interpretation doesn't overturn the Newtonian conception of absolute time and space. Lastly, it could offer additional experimental suggestions which may either strengthen the utility of Lorentz covariance or reveal that it is not a universal principle.

\subsection{Lorentz contraction and clock retardation: the constructive approach}

\noindent
In the paper \textit{"The Ether and the Earth's Atmosphere"} published in Science (1889), FitzGerald proposed that length contraction of a body may occur due to motion relative to an ether:

\begin{quote}
\textit{"We know that electric forces are affected by the motion of electrified bodies relative to the ether and it seems a not improbable supposition that the molecular forces are affected by the motion and that the size of the body alters consequently."}
\end{quote}

\noindent
This was partly motivated by Heaviside's discovery in 1888 that electric fields are contracted in the direction of motion according to Maxwell's equations. To elaborate on FitzGerald's argument, let us consider a stationary configuration in the $xy$-plane of four equal electric charges $q_e$, placed at the vertices of a square with the following coordinates:

\begin{minipage}[t]{0.25\textwidth}
$     \textbf{r}_1 =   \dfrac{1}{2} \begin{pmatrix} - R \\  - R  \end{pmatrix}     $
\end{minipage}
\begin{minipage}[t]{0.25\textwidth}
$      \textbf{r}_2 =   \dfrac{1}{2} \begin{pmatrix}  - R \\  R  \end{pmatrix}           $
\end{minipage}
\begin{minipage}[t]{0.25\textwidth}
$       \textbf{r}_3 =   \dfrac{1}{2} \begin{pmatrix}  R \\  -  R  \end{pmatrix}     $
\end{minipage}
\begin{minipage}[t]{0.25\textwidth}
$     \textbf{r}_4 =   \dfrac{1}{2} \begin{pmatrix}   R \\  R  \end{pmatrix}        $
\end{minipage}

\noindent
A fifth electric charge $  - q_e (1 + 2 \sqrt{2}) / 4 $ is placed at the centre of the square (at the origin). This configuration of five charges is in an electrostatic (albeit unstable) equilibrium, which means that the force acting on each particle vanishes. Let us assume that the system has been accelerated until reaching a steady proper velocity $  \boldsymbol{\beta}  $. If the LT is applied to Coulomb's law, one obtains the EM force between two uniformly moving point charges $ e_j $ and $ e_k $ with the same velocity:
$$    \textbf{F}_{jk}   =   k_e  \frac{ e_j   e_k  }{r^2 }  \left(  \frac{ 1 - \beta^2 }{  \left( 1 -     |   \hat{\textbf{r}}  \times  \boldsymbol{\beta}  |^2  \right)^{3/2}  } \right)     \Big(    \hat{\textbf{r}}     +  \boldsymbol{\beta}  \times  (  \boldsymbol{\beta}  \times  \hat{\textbf{r}} )   \Big)      \hskip \textwidth minus \textwidth  \text{(Lorentz transformed Coulomb's law)}  $$
where $ \hat{\textbf{r}}  $ is the unit vector between the two point charges. If the shape of the system is Lorentz contracted in the direction of motion, then the five electric charges are in equilibrium. This result can be generalized: when any equilibrium configuration is accelerated to a proper velocity $ \boldsymbol{\beta}$ and is Lorentz contracted in the direction of motion, then the resulting system will be in equilibrium. Note, however, that in an equilibrium configuration, the forces may also be viewed as being transmitted instantaneously from one particle to the other.

\begin{figure}[h]
\centering
\includegraphics[scale=.3]{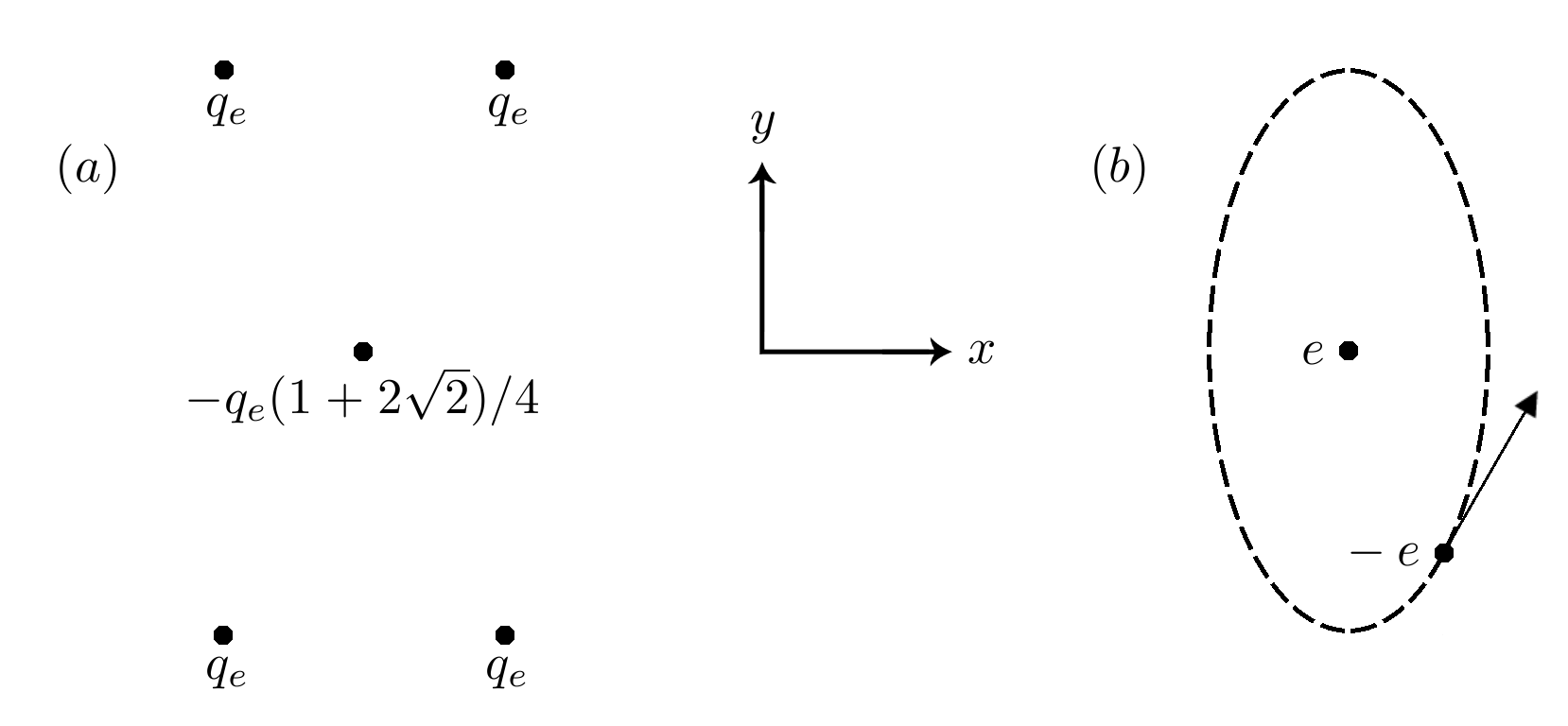}
\caption{The two systems move in the $x$ direction. (a) The five-particle system is Lorentz contracted in the direction of motion. (b) The hydrogen atom, when modeled as an electron orbiting a proton, is Lorentz contracted in the direction of motion and the period of the electron is dilated.}
\label{contraction}
\end{figure}

\noindent
Clock retardation can also coexist with IAAD. In his famous essay \textit{"How to teach special relativity"} published in 1976, Bell pointed out that if we consider an electron orbiting a proton and if we ignore the EM field produced by the electron, then the orbit of the electron is flattened and the period of the electron is dilated by the Lorentz factor when the system attains a certain velocity. In 1986, Bell suggested that certain paradoxes in quantum mechanics may be resolved by reintroducing a PRF.\cite{bell1} He argued that the Lorentzian interpretation is perfectly coherent and that it is the condition of Lorentz covariance which creates difficulties for a realistic interpretation of quantum mechanics.

\noindent
In conclusion, the principle theory does not render redundant the constructive approach. The constructive approach consists of either proving that the physical laws obey Lorentz covariance or showing through other means that a moving body undergoes Lorentz contraction and clock retardation. We mention here some publications in the area of constructivism, showing the actuality of this topic.\cite{brown}-\cite{jarvinen2}

\subsection{A comparison between special relativity theory and Lorentzian instantaneous action at a distance theories}

\noindent
In SRT, all IRFs are equivalent to one another and there are two universal speed limits:
\begin{enumerate}
\item \textit{Inertial constraint:} a body cannot travel faster than the speed of light.
\item \textit{Strong locality:} causal influences cannot travel faster than the speed of light.
\end{enumerate}
In SRT, the inertial constraint is a consequence of the relativistic mass formula $m = m_0 \gamma $, whereas strong locality requires an auxiliary assumption, namely causality. In an IAADT, the inertial constraint still holds, but strong locality is violated. The IAAD forces are distinct from radiative phenomena, which are constrained by the speed limit. To measure the elapsed time between two events, a convention for synchronizing clocks is used. According to the Einstein synchronization method, a light signal is sent at time $ t _1$ from clock $1$ to clock $2$ and immediately back. If its arrival time at clock $1$ is $ t _2$, then clock $2$ is set in such a way that the time $ t_3$ of signal reflection is: $  t_3  =  (t_1 + t_2 ) /2  $. Within an IAADT, there should be an absolute synchronization method that gives rise to the concept of absolute simultaneity. When the absolute synchronization method is used, the one-way speed of light is anisotropic and depends on the absolute speed of the reference frame in which it is measured. Furthermore, Lorentz contraction and clock retardation occur relative to the PRF and entail the isotropy of the two-way speed of light. The clock reading obeys the law $t = T / \gamma $, so the clock reading does not represent the real time $ T $, which is absolute. In 1977, Mansouri and Sexl formulated a test theory for SRT,\cite{mansouri} which describes a transformation between a PRF (with space-time coordinates $X$ and $T$) and an IRF (with a velocity $v$ with respect to the PRF and space-time coordinates $x$ and $t$). The transformation includes three arbitrary functions $ a [v] $, $b[v]$ and $ \epsilon [v] $ of the velocity $v$:
$$    \begin{pmatrix}  t  \\  x   \end{pmatrix} =   
 \begin{pmatrix}   a    -  \epsilon   b   v     &       \epsilon    b     \\    - bv  &       b   \end{pmatrix}    \begin{pmatrix}    T  \\   X    \end{pmatrix}   \hskip \textwidth minus \textwidth   \text{(Mansouri-Sexl transformation)}   $$
In the Mansouri-Sexl test theory, the LT assumes $ 1 / a  =  b  =  \gamma   $ and $ \epsilon  =  - v/c^2  $:
$$    \begin{pmatrix}  t  \\  x   \end{pmatrix} =   \gamma 
 \begin{pmatrix}    1       &       -     v  / c^2     \\    - v   &       1    \end{pmatrix}    \begin{pmatrix}    T  \\   X    \end{pmatrix}   \hskip \textwidth minus \textwidth   \text{(Lorentz transformation)}   $$
Tangherlini investigated a transformation that maintains absolute simultaneity.\cite{tangherlini} The Tangherlini transformation (TT) is represented in the Mansouri-Sexl test theory as $1 / a =  b = \gamma  $ and $ \epsilon = 0  $:
$$    \begin{pmatrix}  t  \\  x   \end{pmatrix} =   
 \begin{pmatrix}   1 / \gamma       &      0    \\    -  \gamma  v  &       \gamma    \end{pmatrix}    \begin{pmatrix}    T  \\   X    \end{pmatrix}   \hskip \textwidth minus \textwidth   \text{(Tangherlini transformation)}   $$
Let us summarize the properties of these two competing paradigms in a single table:

\noindent
\begin{tabular}{ | p{4cm} | p{5cm} | p{5cm} |}
    \hline
     &   SRT: & IAADT: \\ \hline
  Reference frames:   & All IRFs are relative, so there is no PRF.   &   There is a PRF, which is undetectable in conventional optical experiments.  \\ \hline
     Inertial constraint: & \checkmark &  \checkmark    \\ \hline
      Strong locality:            &       \checkmark       &       \text{\sffamily X}               \\  \hline
     Synchronization method: & Einstein synchronization. & Absolute synchronization.  \\ \hline
Isotropy in the one-way speed of light:  &   \checkmark   &   \text{\sffamily X}              \\  \hline
Isotropy in the two-way speed of light:  &   \checkmark   &    \checkmark   \\  \hline
  Lorentz contraction:   & Lengths contract relative to a stationary observer. &  A measuring rod contracts due to motion relative to the PRF according to $L = L_0  / \gamma  $. \\ \hline
     Clock retardation: & Clocks slow relative to a stationary observer. Time is relative: time dilation between IRFs is reciprocal. & A moving clock slows due to motion relative to the PRF according to $ t = T / \gamma $. The real time $ T $ is absolute.   \\ \hline
 Transformation: &   LT.  &  TT.  \\  \hline 
    \end{tabular}

\newpage
\section{Arguments in favour of instantaneous action at a distance}

\noindent
IAAD solves many of the problems associated with a system of point charges: the IAAD forces between point particles are well-defined, there is no diverging energy problem for instantaneous interparticle interactions, there is no infinite regression and IAAD does not pose an immediate threat to causality. It is unlikely that these issues can be settled through mere theoretical arguments, so let us point out some empirically based arguments for reconsidering IAAD:

\begin{enumerate}
\item The dipole anisotropies in the CMB, the galactic red shifts and the muon flux appear to favour the existence of a PRF.
\item Bell's inequality is violated for space-like separated entangled particles.
\item Some experiments call into question the applicability of the standard retardation constraint to all components of the EM field.
\end{enumerate}

\noindent
\textbf{Argument 1:} By measuring the spectrum of the CMB \cite{conklin}-\cite{smoot} and the galaxies \cite{vaucouleurs,rubin} in different directions, our actual speed relative to the CMB and the universe at large has been estimated to be $\sim   0.1  \%$ of the speed of light. Similarly, a dipole anisotropy in the cosmic-ray muon flux can be detected using a cosmic-ray telescope.\cite{monstein} Although this evidence is circumstantial, it may be argued that these dipole anisotropies favour the existence of a PRF.\cite{prokhovnik} Indeed, if the relativity principle is a perfectly valid postulate (and all the IRFs are equivalent), then we would expect that there is no means of discerning whether or not we are in absolute motion. And yet, on the face of it, cosmological observations cast doubt on that supposition. The relativity of simultaneity fundamentally hinges on the relativity principle, so the existence of a cosmic rest frame may suggest that the concept of simultaneity is absolute and that there is a cosmological time arrow. If the hypothetical PRF is the reference frame in which the universe at large is isotropic, then the dipole anisotropies may provide a means of absolute synchronization.

\noindent
\textbf{Argument 2:} The (invariant) spacetime interval between two events $ E_1 = ( \textbf{r}_1,t_1)$ and $ E_2  =  ( \textbf{r}_2 , t_2 )$ is $ |  \textbf{r}_2 - \textbf{r}_1  |^2  - c^2  (t_2 - t_1 )^2 $. There are three types of spacetime intervals:
\begin{itemize}
\item[ ] $   |  \textbf{r}_2 -  \textbf{r}_1  |^2  - c^2  (t_2 - t_1 )^2   < 0   \hskip \textwidth minus \textwidth   \text{(Time-like interval)}   $
\item[ ] $   |   \textbf{r}_2 -  \textbf{r}_1  |^2  - c^2  (t_2 - t_1 )^2    = 0  \hskip \textwidth minus \textwidth   \text{(Light-like interval)}   $
\item[ ] $   |   \textbf{r}_2 -  \textbf{r}_1  |^2  - c^2  (t_2 - t_1 )^2    > 0   \hskip \textwidth minus \textwidth   \text{(Space-like interval)}   $
\end{itemize}

\noindent
The past light cone of an event $ \mathcal{O} $ contains all the events from which one can reach $ \mathcal{O}$ by future directed timelike or lightlike trajectories. In a strongly local theory, any correlation between two space-like separated events $ E_1 =(  \textbf{r}_1,t_1) $ and $ E_2 =(  \textbf{r}_2 ,t_2) $ arises from each of them being correlated with events within the intersection of their past light cones:
$$   P [ E_1 , E_2    |  L_1 , L_2 , \lambda  ]  =   P [ E_1   |  L_1  , \lambda  ]      P [  E_2   |  L_2 , \lambda  ]        \hskip \textwidth minus \textwidth   \text{(Strong locality)}   $$
where $ L_1 $ is the set of events in the past light cone of $ E_1 $, but not $E_2$, $L_2$ is the set of events in the past light cone of $E_2$, but not $ E_1 $, and $  \lambda $ is the intersection of the past light cones of $E_1$ and $E_2$. If $ E_1 $ and $ E_2 $ are space-like separated, then $ E_1 $ and $ E_2 $ are independent of each other; causal influences cannot travel faster than the speed of light. Strong locality is sometimes referred to as "factorizability" because it entails the factorizability of the probability function for space-like separated events.\cite{alford}

\begin{figure}[h]
\centering
\includegraphics[scale=0.6]{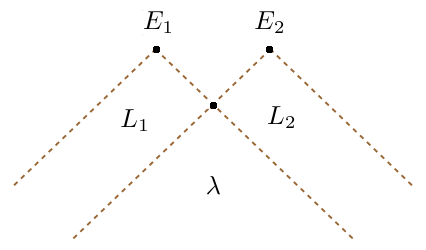}
\caption{Strong locality states that any correlation between two space-like separated events arises from each of them being correlated with events within the intersection of their past light cones.}
\end{figure}

\noindent
Suppose there are two space-like separated particles $1$ and $2$. Furthermore, there are three two-valued properties $A$, $B$ and $C$. For particle $j \in \{  1,2  \}$ and property $X \in \{  A, B, C \}$, we define a stochastic variable $X_j$:
$$ \begin{cases}
    X_j = 0       & \quad \text{if particle } j  \text{ does not have property }X\\
    X_j = 1  & \quad  \text{if particle } j  \text{ has property }X \\
  \end{cases}    \hskip \textwidth minus \textwidth   \text{ } $$
Furthermore, for any property $X \in \{  A, B, C \}$, the following correlation holds:
$$  P [ X_1  \neq  X_2  ]  = 1    \hskip \textwidth minus \textwidth  \text{(Entanglement)}     $$

\begin{figure}[h]
\centering
\includegraphics[scale=0.25]{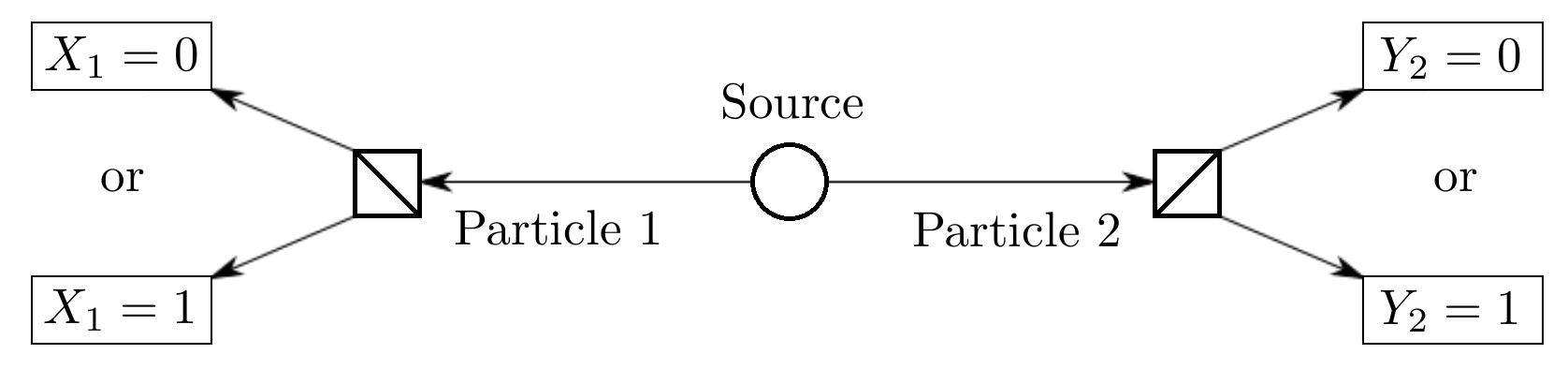}
\caption{An illustration of a Bell test experiment for the properties $ X \in \{ A,B,C \}$ and $ Y \in  \{ A,B,C \} $.}
\end{figure}

\noindent
Bell's inequality refers to the correlation between measurement outcomes of different properties.\cite{bell2} This inequality is a corollary of the following logical implication:
$$  (  B_2 =  A_1  \land  A_1 = C_2  \land     C_2 = B_1  )  \Rightarrow     B_2 =  B_1   \hskip \textwidth minus \textwidth  \text{ }  $$
The contrapositive of this implication is:
$$    B_2 \neq  B_1  \Rightarrow  (  B_2 \neq  A_1  \lor  A_1 \neq  C_2  \lor     C_2 \neq B_1  )     \hskip \textwidth minus \textwidth  \text{ }  $$
which can be written as a statement about probabilities:
$$  P [ B_2  \neq B_1   ]  \leq  P[  B_2 \neq  A_1 ] + P [  A_1 \neq  C_2 ]  + P[  C_2 \neq B_1 ] \hskip \textwidth minus \textwidth  \text{ } $$
From the entanglement of the particles, it follows that:
$$ 1  \leq  P[  B_2 \neq  A_1 ] + P [  A_1 \neq  C_2 ]  + P[  C_2 \neq B_1 ]     \hskip \textwidth minus \textwidth  \text{(Bell's inequality)}   $$
The experiments show that Bell's inequality is violated for space-like separated entangled particles,\cite{hensen} which may suggest that the measurement setting for one particle affects the outcome of the measurement on the other particle even when the two particles are space-like separated. One should of course be cautious in suggesting IAAD as an explanation for Bell test experiments, because several assumptions are left implicit (such as counterfactual definiteness, freedom of choice and causality). In light of SRT, weaker versions of locality have also been investigated. In particular, signal locality states that superluminal signaling is prohibited. The difference between a causal influence and a signal is that the latter requires a controllable means of transferring information. It is widely believed that the violation of Bell's inequality for space-like separated events cannot be exploited to allow superluminal signaling, due to the no-signaling theorem.\cite{eberhard,ghirardi}

\noindent
\textbf{Argument 3:} The contemporary experiments related to the propagation speed of the different components of the EM field can be divided into two categories:

\begin{enumerate}[-]
\item \textit{Experiments to measure the propagation speed of the velocity-dependent electric field:} Some preliminary attempts have been made to measure the propagation speed of the velocity-dependent electric field. It appears that the velocity-dependent electric field generated by an electron beam moving uniformly for a finite time is rigidly carried by the beam itself, contrary to the standard retardation constraint.\cite{pizzella} The response that was expected from a completely retarded electric field was orders of magnitude smaller than what was actually observed. An alternative interpretation of these experimental results in terms of the acceleration-dependent part of the retarded EM field has also been investigated.\cite{shabad,pizzella1}

\begin{figure}[h]
\centering
\includegraphics[scale=0.3]{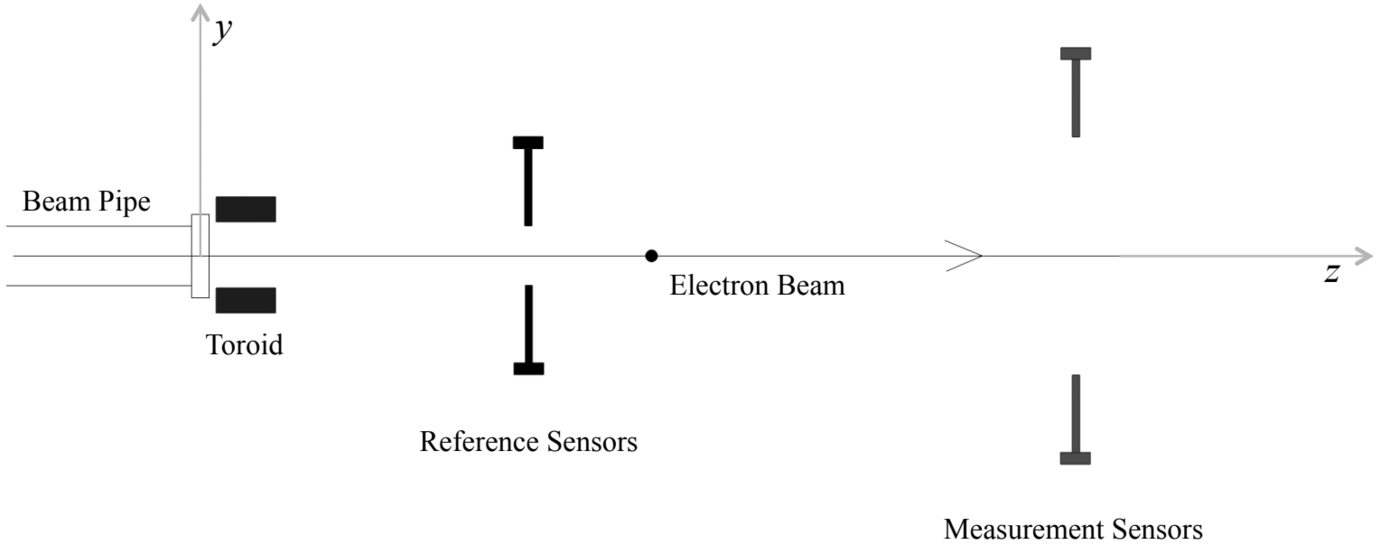}
\caption{The arrangement of an experiment, performed by Pizzella et al., to measure the propagation speed of the velocity-dependent electric field. The electron beam is produced by the beam pipe. The movable measurement sensors are used to measure the electric field at different locations.}
\end{figure}

\item \textit{Experiments to measure the propagation speed of the bound magnetic field:} In classical electrodynamics, the magnetic field produced by a neutral stationary wire $\mathcal{C}$ is composed of a bound magnetic field and a radiative magnetic field:
$$    \textbf{B}   [  \textbf{x}  , t ]  =    \underbrace{  k_m      \oint_\mathcal{C}      \frac{  I  [ t - r/c] }{ r^2 }    d  \boldsymbol{\ell}   \times    \hat{\textbf{r}}  }_\text{Bound magnetic field} +   \underbrace{ k_m  \oint_\mathcal{C}  \frac{  \dot{I} [t - r/c ] }{ c r }           d  \boldsymbol{\ell}   \times    \hat{\textbf{r}}  }_\text{Radiative magnetic field}  \hskip \textwidth minus \textwidth  \text{ }  $$
where $ I $ is the current in the wire, $ \boldsymbol{\ell} $ is a point on the path $ \mathcal{C} $, $ \textbf{r} = \textbf{x}  -  \boldsymbol{\ell} $ is the displacement vector from $ \boldsymbol{\ell} $ to $ \textbf{x} $ and $r = | \textbf{r} |$ is the distance from the source. The bound magnetic field falls off as $  r^{-2} $, while the radiative magnetic field falls off as $ r^{-1} $. Hence, the bound magnetic field dominates in the near region, while the radiative magnetic field dominates in the far region. There is a slight difference between the bound magnetic field and the earlier mentioned velocity-dependent magnetic field. For steady currents, the velocity-dependent magnetic field does not coincide with the Biot-Savart law, but the discrepancy is exactly cancelled out by the field due to the acceleration of the conduction electrons in negotiating the bends. In 1888, Heinrich Hertz provided convincing evidence that the EM field in the far region satisfies the standard retardation constraint, but no experimental attempt was made to separate the effect of different EM field components.\cite{hertz} Nevertheless, Hertz's experimental results gave rise to the idea that all the EM field components propagate at the exact same speed. There are, however, some experiments that call into question the applicability of the standard retardation constraint to the bound magnetic field.\cite{kholmetskii1}-\cite{kholmetskii4} In these experiments, the near and far magnetic fields were investigated by measuring the electromotive force (EMF) at different distances from an emitting antenna. Let the emitting (EA) and receiving (RA) antennas be circular coils both of surface area $ \Delta S $. The following model equation for the magnetic field produced by the EA was used:
$$    \textbf{B}  [ \textbf{x} , t ]  =      k_m      \oint_\text{EA}   \left(   \frac{  [ I ]_v }{ r^2 }   +   \frac{  [ \dot{I}]_c }{ c r }   \right)        d  \boldsymbol{\ell}   \times  \hat{\textbf{r}}   \hskip \textwidth minus \textwidth  \text{(Model equation)}  $$
The square brackets indicate retardation of the enclosed quantity and the subscript indicates the speed of propagation: the enclosed quantity is evaluated at the time $t - r/v$ for the bound magnetic field and at the time $t - r/c$ for the radiative magnetic field. Classical electrodynamics satisfies the standard retardation constraint $v=c$, but the propagation speed is introduced as a free parameter in the model equation. Recall that a magnetic dipole is the limit of a circular loop as the radius of the loop reduces to zero while keeping the magnetic moment constant. In the spherical coordinate system, the model equation associated with an oscillating magnetic dipole is:
$$   \textbf{B}  [ \textbf{x} , t]  =   \underbrace{   k_m  \Delta S    \left(     \cos [ \theta  ]   \frac{     [ I ]_v    }{ r^3 }    \hat{\textbf{r}}              +     \sin [ \theta  ]    \left(  \frac{  [ I ]_v  }{ r^3 }  +   \frac{ [ \dot{I} ]_v   }{ v  r^2 }      \right)      \hat{\boldsymbol{\theta}}         \right)         }_\text{Bound magnetic field}                                   
  +   \underbrace{    k_m  \Delta S      \left(   \cos [ \theta  ]        \frac{   [ \dot{I} ]_c  }{ c r^2  }     \hat{\textbf{r}}    +   \sin [ \theta  ]      \frac{  [ \ddot{I} ]_c  }{ c^2 r }  \hat{\boldsymbol{\theta}}    \right)  }_\text{Radiative magnetic field}      \hskip \textwidth minus \textwidth  \text{ }  $$

\begin{figure}[h]
\centering
\includegraphics[scale=0.35]{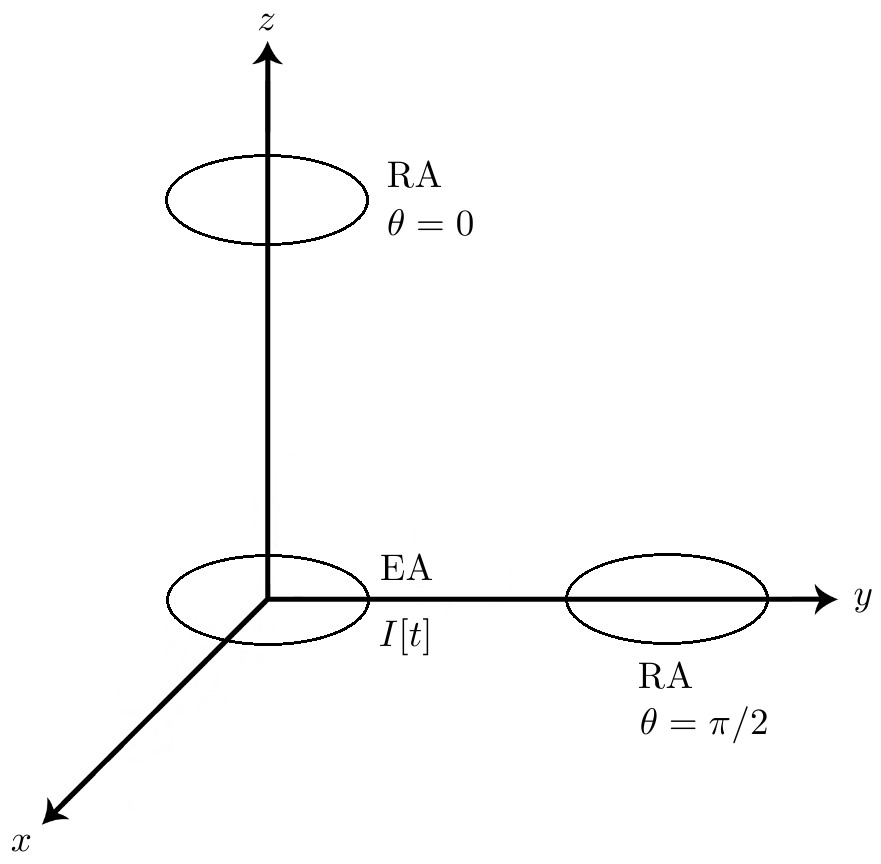}
\caption{The arrangement of the experiments, performed by Kholmetskii et al., to measure the propagation speed of the bound magnetic field. In the coplanar configuration ($ \theta = \pi /2 $), the EA and RA belong to the same plane and in the coaxial configuration ($   \theta = 0  $), the EA and RA share the same axis of symmetry.}
\end{figure}

\noindent
In the coplanar configuration ($ \theta = \pi /2 $), the EA and RA belong to the same plane and in the coaxial configuration ($   \theta = 0  $), the EA and RA share the same axis of symmetry:
$$     \textbf{B}_\text{pl}   [  \textbf{x} , t ]   =    -  k_m \Delta S   \left(  \frac{ [I]_v  }{  r^3}  +   \frac{ [\dot{I}]_v}{ v r^2 }   +  \frac{ [ \ddot{I} ]_c }{c^2  r }  \right)   \hat{\textbf{z}}      \hskip \textwidth minus \textwidth  \text{ }  $$
$$     \textbf{B}_\text{ax}   [  \textbf{x} , t ]   =     k_m \Delta S   \left(  \frac{ [I]_v  }{  r^3}  +   \frac{ [\dot{I}]_c }{ c r^2  }    \right)   \hat{\textbf{z}}      \hskip \textwidth minus \textwidth  \text{ }  $$

\noindent
where the unit vector $\hat{\textbf{z}}$ is perpendicular to the antennas. According to Faraday's law of induction, the EMF in the RA is: $    \epsilon [t]   =  -   d     \Phi_B    / dt         $, where $ \Phi_B $ is the magnetic flux enclosed by the RA. Hence, the EMFs in the coplanar and coaxial configurations are:
$$    \epsilon_\text{pl}   [t]  =    k_m  ( \Delta S )^2   \left(   \frac{  [ \dot{I} ]_v }{ r^3}   +   \frac{ [ \ddot{I}]_v }{v r^2}  + \frac{ [ \dddot{I} ]_c}{c^2 r}    \right)         \hskip \textwidth minus \textwidth  \text{ }  $$
$$    \epsilon_\text{ax} [t]      =  -  k_m  ( \Delta S )^2   \left(   \frac{  [ \dot{I} ]_v }{ r^3 }   +   \frac{ [ \ddot{I}]_c }{c r^2}    \right)         \hskip \textwidth minus \textwidth  \text{ }  $$
If the current in the EA oscillates harmonically at an angular frequency $ \omega $ as $ I[t] = I_0 \cos [ \omega t  ] $, then the EMFs can be written as:
$$   \epsilon_\text{pl} [t]   = \epsilon_0    \left(  - \frac{ \sin [ \omega ( t - r / v ) ] }{ r^3}  -  \frac{ \omega  \cos [  \omega (t - r/ v)   ]  }{ v r^2 }   +  \frac{ \omega^2   \sin [  \omega  (  t-  r  / c ) ]  }{ c^2  r }  \right)     \hskip \textwidth minus \textwidth  \text{ }  $$
$$   \epsilon_\text{ax} [t]   = \epsilon_0    \left(   \frac{ \sin [ \omega ( t -  r / v ) ] }{ r^3  } +  \frac{ \omega  \cos [  \omega (t - r / c )   ]  }{ c r^2 }    \right)     \hskip \textwidth minus \textwidth  \text{ }  $$

\noindent
The size $R_n$ of the near region is proportional to the wavelength of the emitted radiation: $ R_n =  c  /  \omega $. The reference signal $\epsilon_{\text{ref}} [t]$ can be determined by measuring the EMF at large distances where only the radiative contribution remains. This function can be extrapolated back to the near region. The first experiments \cite{kholmetskii1,kholmetskii2} used the zero crossing method to determine the parameter $v$ in a coplanar configuration. The time difference $\Delta  t  [R] = t_2 - t_1$ is the difference between the zero crossing point of the total signal ($ \epsilon  [ t_2  ] = 0$) and the zero crossing point of the reference signal ($ \epsilon_{\text{ref}} [ t_1 ]  = 0 $). The measured results for $ \Delta t$ in comparison with the numerical predictions, are shown in Figure \ref{speed-bound-field}. A striking coincidence with an instantaneous spreading velocity $ v = \infty $ was found within the near region. More advanced experiments \cite{kholmetskii3,kholmetskii4} by the same group investigated the propagation speed of the bound magnetic field in the near and far region and applied a decomposition procedure to the entire time interval of the signal. The experiments are once again at odds with the standard retardation constraint and show an anomalously small retardation of the bound magnetic field in approximately half the near region, whereas the propagation speed of the bound magnetic field in the far region tends toward the speed of light. The fact that the propagation speed of the bound magnetic field is not constant and that the region in which the bound magnetic field manifests anomalously small retardation scales as the near region, suggests that the model equation that was obtained within the framework of classical electrodynamics requires a substantial revision, which is outside the scope of this paper.

\end{enumerate}

\begin{figure}[h]
\centering
\includegraphics[scale=0.4]{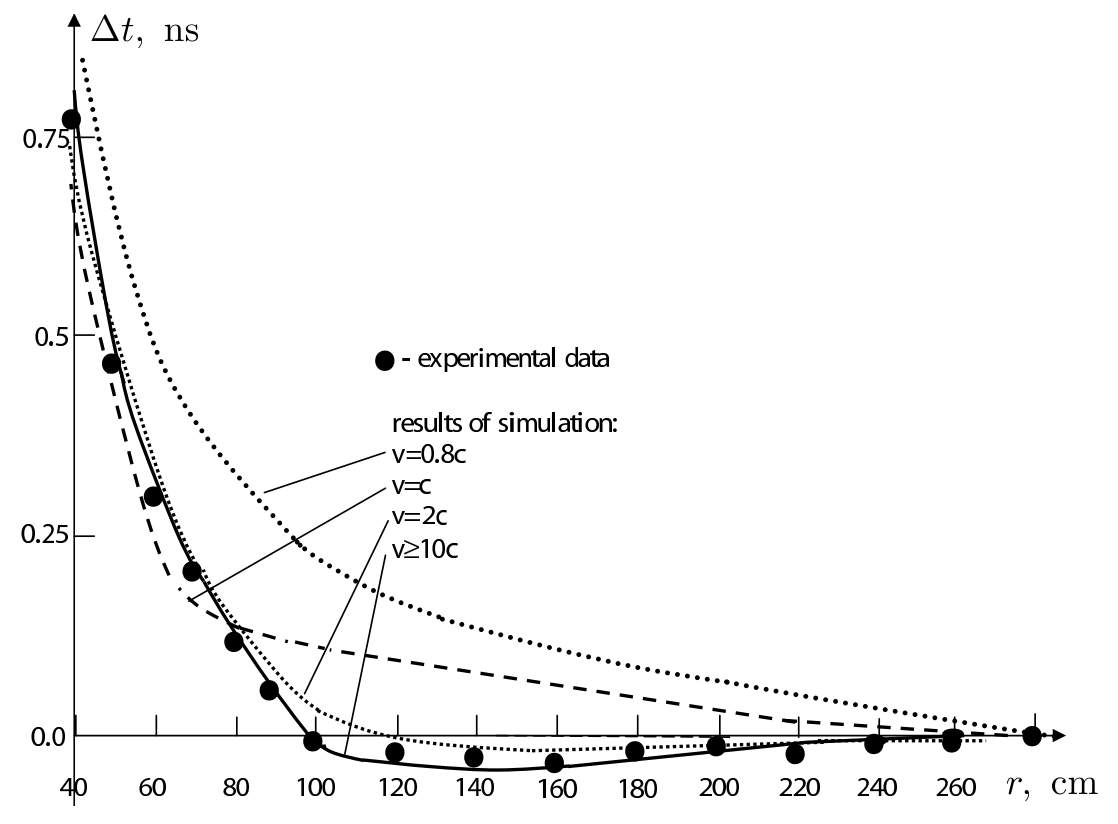}
\caption{The zero crossing method was used to determine the propagation speed of the bound magnetic field in a coplanar configuration. Dot lines illustrate numerical predictions of $\Delta t $ for the retardation conditions $v= 0.8c$, $v = c$ and $v = 2c$. The limit case $v \geq 10 c  $ is plotted as a continuous line. Experimental data are represented by black circles.\cite{kholmetskii2}}
\label{speed-bound-field}
\end{figure}

\newpage
\section{Field-free interaction Lagrangian}

\noindent
Now that we have laid out several arguments for reconsidering IAAD, we are in a position to develop IAADTs of electrodynamics. The field-free Lagrangian can be written as:
$$    \mathcal{L} =   \sum_j   \left(   -   m_{0j}  c^2  \sqrt{1 -   \beta_j^2 }   +   \frac{1}{2}   \sum_{ j \neq k  }   \mathcal{L}_{jk}       \right)       \hskip \textwidth minus \textwidth  \text{ }  $$
where $ m_{0j} $ and $ \boldsymbol{\beta}_j $ are, respectively, the rest mass and the proper velocity of particle $j$ and $\mathcal{L}_{jk}$ is the field-free interaction Lagrangian describing the interaction between particles $ j $ and $ k $. The motion of particle $\ell$ is determined by the Euler-Lagrange equation:
$$     \frac{d}{dt}  \frac{\partial \mathcal{L}}{\partial \dot{\textbf{r}}_\ell} =  \frac{\partial \mathcal{L} }{ \partial \textbf{r}_\ell }    \hskip \textwidth minus \textwidth  \text{ }  $$
The force acting on particle $j$ due to particle $k$ is given by:
$$   \textbf{F}_{jk}  =  \frac{  \partial    \mathcal{L}_{jk}  }{  \partial   \textbf{r}_j   }  -   \frac{d}{ c dt}  \frac{ \partial \mathcal{L}_{jk}  }{ \partial  \boldsymbol{\beta}_j }      \hskip \textwidth minus \textwidth  \text{ }  $$
We will explore two different field-free interaction Lagrangians and demonstrate their compatibility with some basic features of classical electrodynamics.

\subsection{Frejlak's field-free electrodynamics}

\noindent
First, we wish to present the following field-free interaction Lagrangian that was first discovered by Wojciech Frejlak in 1988:\cite{frejlak}
$$   \mathcal{L}^{(F)}_{jk} =   -  k_e  \frac{ e_j  e_k  }{  r }   \frac{1 - \boldsymbol{\beta}_j   \cdot  \boldsymbol{\beta}_k  }{  \sqrt{1-  (  \hat{\textbf{r}}   \times  \boldsymbol{\beta}_j  )  \cdot  ( \hat{\textbf{r}}   \times  \boldsymbol{\beta}_k  )  } }     \hskip \textwidth minus \textwidth  \text{(Frejlak's field-free Lagrangian)}  $$
where $ \textbf{r} = \textbf{r}_{jk}  =  \textbf{r}_j - \textbf{r}_k $, $ r =   |  \textbf{r} | $ and $ \hat{\textbf{r}} =  \textbf{r} /r $. In the low-velocity limit, this Lagrangian reduces to Darwin's field-free Lagrangian $  \mathcal{L}^{(F,2)}_{jk} $, which was discovered by Charles Galton Darwin in 1920:\cite{darwin}
$$  \mathcal{L}^{(F,2)}_{jk}     =     - k_e  \frac{   e_j    e_k   }{r}    \left(  1  -  \frac{  \boldsymbol{\beta}_j  \cdot  \boldsymbol{\beta}_k }{2}  -  \frac{   ( \hat{\textbf{r}}  \cdot   \boldsymbol{\beta}_j  ) (   \hat{\textbf{r}}  \cdot   \boldsymbol{\beta}_k     )  }{2}   \right)      \hskip \textwidth minus \textwidth  \text{(Darwin's field-free Lagrangian)}  $$
where we have kept terms up to the second order in $1/c$ (hence the superscript $2$). Frejlak's field-free Lagrangian has a simple analytical form and cures some deficiencies of Darwin's field-free Lagrangian in the high-velocity regime.

\subsection{Lorenz's field-free electrodynamics}

\noindent
Let us now assume a field-free Lagrangian that takes the following form:
$$    \mathcal{L} =   \sum_j   \left(   -   m_{0j}  c^2  \sqrt{1 -   \beta_j^2 }   -  \frac{ e_j }{2}   \sum_{k \neq j }  ( \phi_k [ \textbf{r}_j ,t   ]   -   \dot{\textbf{r}}_j  \cdot   \textbf{A}_k [ \textbf{r}_j  ,t  ]    )       \right)       \hskip \textwidth minus \textwidth  \text{ }  $$
where the EM potentials $ \phi $ and $\textbf{A} $ of a uniformly moving point charge are used in a particular gauge. If we impose the Lorenz gauge condition, then the field-free interaction Lagrangian is:
$$       \mathcal{L}^{(L)}_{jk}      =  -    k_e  \frac{   e_j    e_k  }{ 2 r  }   \left(     \frac{ 1 - \boldsymbol{\beta}_j  \cdot  \boldsymbol{\beta}_k }{  \sqrt{   1-    | \hat{\textbf{r}} \times  \boldsymbol{\beta}_j   |^2     } }   +   \frac{ 1 - \boldsymbol{\beta}_j  \cdot   \boldsymbol{\beta}_k }{  \sqrt{   1-     | \hat{\textbf{r}} \times   \boldsymbol{\beta}_k  |^2     } }        \right)           \hskip \textwidth minus \textwidth  \text{(Lorenz's field-free Lagrangian)}  $$
In the low-velocity limit, this Lagrangian reduces to:
$$   \mathcal{L}^{(L,2)}_{jk}    =     - k_e  \frac{  e_j   e_k }{  r  }     \left(      1 - \boldsymbol{\beta}_j  \cdot  \boldsymbol{\beta}_k       + 
    \frac{ \beta_j^2  +  \beta_k^2  }{ 4 }     -  \frac{     ( \hat{\textbf{r}} \cdot  \boldsymbol{\beta}_j  )^2 +    (  \hat{\textbf{r}} \cdot  \boldsymbol{\beta}_k  )^2    }{ 4 }     \right)    \hskip \textwidth minus \textwidth  \text{ }  $$
where we have, once again, only kept terms up to the second order in $1/c$. This type of electrodynamics may be referred to as Lorenz's field-free electrodynamics, since the Lorenz gauge was used to construct the field-free interaction Lagrangian. To the best of the authors' knowledge, it has not been presented before.

\newpage
\section{Compatibility of classical and field-free electrodynamics}

\noindent
The force between two stationary charges is described by Coulomb's inverse-square law, which means that electrostatics is embodied in Frejlak's and Lorenz's field-free electrodynamics. Let us derive the following basic laws of classical electrodynamics:

\begin{enumerate}
\item Lorentz transformed Coulomb's law.
\item Biot-Savart force law.
\item Faraday's law of induction.
\end{enumerate}

\subsection{Lorentz transformed Coulomb's law}

\noindent
Consider two uniformly moving point charges $1$ and $2$ with the same velocity: $   \boldsymbol{\beta}_1 =   \boldsymbol{\beta}_2 =   \boldsymbol{\beta} $. The resulting force (in both Frejlak's and Lorenz's field-free electrodynamics) is:
$$     \textbf{F}_{12}      =     \frac{ \partial   \mathcal{L}_{12}   }{  \partial  \textbf{r}_1  }  -  \underbrace{  \frac{ d}{ cdt}   \frac{ \partial  \mathcal{L}_{12}  }{  \partial  \boldsymbol{\beta}_1  }    }_{ = 0 }     =        k_e  \frac{   e_1    e_2   }{  r^2 }   \left( \frac{    1 -   \beta^2     }{     (  1   -   |  \hat{\textbf{r}}  \times   \boldsymbol{\beta} |^2      )^{3/2}        }      \right)      \Big(    \hat{\textbf{r}}   +    \boldsymbol{\beta}  \times (  \boldsymbol{\beta}  \times   \hat{\textbf{r}}   )    \Big)       \hskip \textwidth minus \textwidth  \text{ }  $$

\begin{figure}[h]
\centering
\includegraphics[scale=.2]{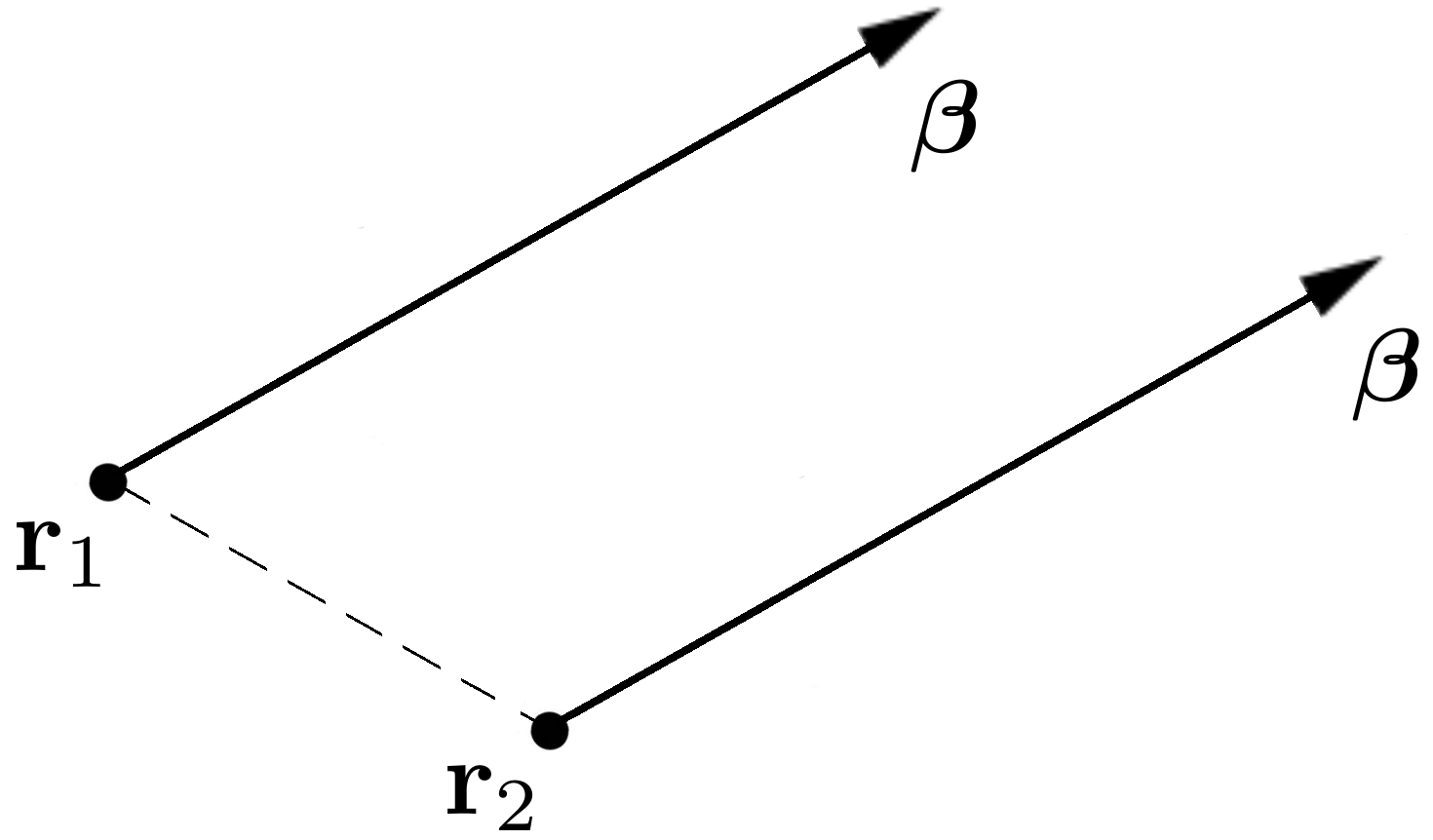}
\caption{Two uniformly moving point charges with the same velocity. The force between the two charges is given by the Lorentz transformed Coulomb's law.}
\end{figure}

\subsection{Biot-Savart force law}\label{biot-savart}

\noindent
Let $ \partial  \Omega_1  $ and $ \partial \Omega_2 $ be closed circuits that current elements $1$ and $2$, respectively belong to. It is assumed that a current element consists of infinitesimal positive and negative charges, $d  e_+ $ and $ d e_-   = - d  e_+ $. Hence, in order to compute the force between the current elements $1$ and $2$, one must add four components: the $++$, $+-$, $-+$ and $--$ interactions. Using $ I_1  d \boldsymbol{\ell}_1  =   d  e_+   ( \dot{\textbf{r}}_{1+}  - \dot{\textbf{r}}_{ 1 - }  ) $ and $ I_2 d \boldsymbol{\ell}_2 = d  e_+ ( \dot{\textbf{r}}_{2+} - \dot{\textbf{r}}_{2-} ) $, we will evaluate the force between two current elements in Frejlak's and Lorenz's field-free electrodynamics.

\begin{figure}[h]
\centering
\includegraphics[scale=.4]{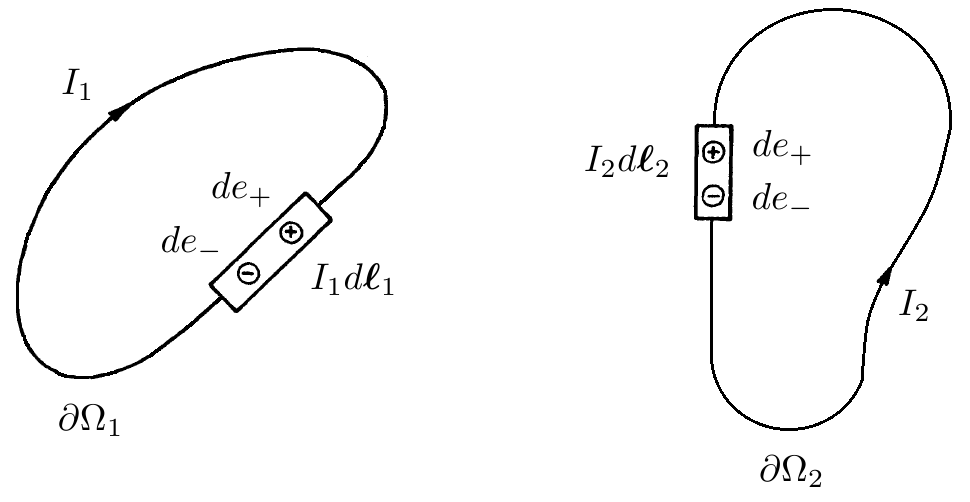}
\caption{In order to obtain the force between two current elements, one must add four components: the $++$, $+-$, $-+$ and $--$ interactions.}
\end{figure}

\subsubsection{Biot-Savart force law in Frejlak's field-free electrodynamics}

\noindent
Using the following identities:

\begin{minipage}[t]{0.3\textwidth}
$    \dfrac{ dr}{dt} =  \hat{\textbf{r}}  \cdot  \dot{\textbf{r}}        $
\end{minipage}
\begin{minipage}[t]{0.7\textwidth}
$    \dfrac{ d  \hat{\textbf{r}}  }{  d t }   =  \dfrac{ \dot{\textbf{r}}   - \hat{\textbf{r}} ( \hat{\textbf{r}} \cdot  \dot{\textbf{r}}  )  }{r}      \hskip \textwidth minus \textwidth   \text{ }  $
\end{minipage}

\noindent
it follows that:
$$    \frac{ \partial  \mathcal{L}^{(F,2)}_{jk}  }{ \partial   \textbf{r}_j   }     =    k_e   e_j   e_k  \frac{ \hat{\textbf{r}} }{ r^2  }   \left(   1    -  \frac{1}{2}   \boldsymbol{\beta}_j  \cdot \boldsymbol{\beta}_k   -    \frac{3}{2}   ( \hat{\textbf{r}}  \cdot   \boldsymbol{\beta}_j  )(  \hat{\textbf{r}} \cdot \boldsymbol{\beta}_k  )       \right)      +   k_e   e_j   e_k \frac{1}{ 2 r^2}  \left(   \boldsymbol{\beta}_j  ( \hat{\textbf{r}} \cdot  \boldsymbol{\beta}_k  )  +     \boldsymbol{\beta}_k   (  \hat{\textbf{r}}  \cdot \boldsymbol{\beta}_j  )  \right)       \hskip \textwidth minus \textwidth  \text{ }  $$
\begin{flalign}\notag
-  \frac{d}{cdt}  \frac{ \partial \mathcal{L}^{(F,2)}_{jk}}{  \partial  \boldsymbol{\beta}_j }   &  =  - k_e   e_j   e_k   \frac{d}{cdt}  \left(   \frac{ \boldsymbol{\beta}_k  + \hat{\textbf{r}}   (  \hat{\textbf{r}}  \cdot  \boldsymbol{\beta}_k   ) }{2r}  \right)  =   &&  \\\nonumber
&  =          -    k_e  e_j    e_k  \frac{1}{2c r^2 }    \left(      \dot{\boldsymbol{\beta}}_k  r  -    \boldsymbol{\beta}_k   (  \hat{\textbf{r}} \cdot   \dot{\textbf{r}}  )      
  -     3   \hat{\textbf{r}}  (  \hat{\textbf{r}}  \cdot  \boldsymbol{\beta}_k  )  (  \hat{\textbf{r}}  \cdot  \dot{\textbf{r}} )     +   
    \dot{\textbf{r}}     ( \hat{\textbf{r}}  \cdot   \boldsymbol{\beta}_k  )    +   \hat{\textbf{r}}  (    \dot{\textbf{r}}  \cdot  \boldsymbol{\beta}_k )   +  \textbf{r} (  \hat{\textbf{r}}  \cdot  \dot{\boldsymbol{\beta}}_k   )      \right)  &&
\end{flalign}

\noindent
Hence, the force between two current elements is given by the Biot-Savart force law (Appendix \ref{magnetostatics-frejlak}):
$$   \textbf{F}^{(F,2)}_{12}    =    k_m    I_1  I_2   \frac{1}{ r^2 }   d \boldsymbol{\ell}_1  \times ( d  \boldsymbol{\ell}_2  \times  \hat{\textbf{r}}  )      \hskip \textwidth minus \textwidth  \text{ }  $$
which means that magnetostatics is embodied in Frejlak's field-free electrodynamics.

\subsubsection{Biot-Savart force law in Lorenz's field-free electrodynamics}

\noindent
In the low-velocity limit, the interparticle force corresponding to Lorenz's field-free Lagrangian is:
$$       \textbf{F}^{(L,2)}_{jk}    =   \frac{ \partial  \mathcal{L}^{(L,2)}_{jk}}{ \partial  \textbf{r}_j  }     - \frac{d}{cdt}  \frac{ \partial   \mathcal{L}^{(L,2)}_{jk} }{ \partial  \boldsymbol{\beta}_j  }           \hskip \textwidth minus \textwidth  \text{ }  $$
where:
\begin{flalign}\notag
  \frac{ \partial  \mathcal{L}^{(L,2)}_{jk}}{ \partial  \textbf{r}_j  } & =     k_e   e_j  e_k  \frac{ \hat{\textbf{r}} }{ r^2  }   \left(      1 - \boldsymbol{\beta}_j  \cdot  \boldsymbol{\beta}_k       + 
    \frac{ \beta_j^2  +  \beta_k^2  }{4}   -  \frac{  3}{4} (  ( \hat{\textbf{r}} \cdot  \boldsymbol{\beta}_j  )^2 +    (  \hat{\textbf{r}} \cdot  \boldsymbol{\beta}_k  )^2     )        \right)  + && \\\nonumber
& +   k_e    e_j    e_k  \frac{1}{ 2 r^2 }   \left(      \boldsymbol{\beta}_j   ( \hat{\textbf{r}}  \cdot  \boldsymbol{\beta}_j ) +  \boldsymbol{\beta}_k  (  \hat{\textbf{r}}  \cdot  \boldsymbol{\beta}_k  )       \right)    && 
\end{flalign}
\begin{flalign}\notag
 - \frac{d}{cdt}  \frac{ \partial   \mathcal{L}^{(L,2)}_{jk} }{ \partial  \boldsymbol{\beta}_j  }   & =   -   k_e  e_j   e_k   \frac{d}{cdt}  \left(  \frac{1}{r}  \left(  \boldsymbol{\beta}_k    -   \frac{  \boldsymbol{\beta}_j  }{2}     +   \frac{   \hat{\textbf{r}}  (  \hat{\textbf{r}}  \cdot  \boldsymbol{\beta}_j   )  }{2}   \right)       \right)     =    &&  \\\nonumber
&   =    k_e    e_j     e_k   \frac{  ( \hat{\textbf{r}}  \cdot   \dot{\textbf{r}} )  }{c  r^2 }  \left(   \boldsymbol{\beta}_k  -  \frac{  \boldsymbol{\beta}_j }{2}   +  \frac{  3  \hat{\textbf{r}} ( \hat{\textbf{r}} \cdot   \boldsymbol{\beta}_j  )  }{2}  \right)   +  &&  \\\nonumber
&  +    k_e   e_j    e_k   \frac{1}{ cr }     \left(    \frac{  \dot{\boldsymbol{\beta}}_j }{2}  - \dot{\boldsymbol{\beta}}_k   -    \frac{  \hat{\textbf{r}}  (  \hat{\textbf{r}}   \cdot \dot{\boldsymbol{\beta}}_j  )  }{ 2  }  -   \frac{   \hat{\textbf{r}}  (    \dot{\textbf{r}} \cdot \boldsymbol{\beta}_j  )  }{ 2 r  }
   -    \frac{    \dot{\textbf{r}}  ( \hat{\textbf{r}}   \cdot    \boldsymbol{\beta}_j    )  }{ 2 r  }   \right)        &&  
\end{flalign}
The force between two current elements is (Appendix \ref{magnetostatics-lorenz}):
$$     \textbf{F}^{(L,2)}_{12}    =    -   k_m    I_1  I_2  \frac{  \hat{\textbf{r}} }{  2 r^2 }         ( d  \boldsymbol{\ell}_1  \cdot d \boldsymbol{\ell}_2 )      +
 k_m    I_1  I_2   \frac{  3  d  \boldsymbol{\ell}_2  }{2 r^2}        (      \hat{\textbf{r}}   \cdot   d  \boldsymbol{\ell}_1  )   +    
k_m    I_1  I_2   \frac{   d  \boldsymbol{\ell}_1  }{ 2  r^2}        (    \hat{\textbf{r}}  \cdot   d  \boldsymbol{\ell}_2   )    
 -    k_m     I_1 I_2  \dfrac{ 3   \hat{\textbf{r}}    }{  2   r^2  }  (   \hat{\textbf{r}}   \cdot  d  \boldsymbol{\ell}_1      )  (   \hat{\textbf{r}}   \cdot  d  \boldsymbol{\ell}_2      )           \hskip \textwidth minus \textwidth  \text{ }  $$
which coincides with the Biot-Savart force law when applied to closed circuits (Appendix \ref{magnetostatics-lorenz}):
$$  \textbf{F}^{(L,2)}_{ 1 , \partial  \Omega_2 }   =    k_m   I_1  d  \boldsymbol{\ell}_1  \times     \oint_{ \partial   \Omega_2  }     I_2   \frac{ d  \boldsymbol{\ell}_2  \times  \hat{\textbf{r}}  }{  r^2  }     \hskip \textwidth minus \textwidth  \text{ }     $$
where $ \textbf{F}^{(L,2)}_{1,  \partial  \Omega_2}  $ is the force that circuit $ \partial \Omega_2 $ exerts on current element $1$. We conclude that magnetostatics is embodied in Lorenz's field-free electrodynamics. In Appendix \ref{generalized-lagrangian}, we determine the conditions under which a generalized field-free interaction Lagrangian shows consistency with the Lorentz transformed Coulomb's law and the Biot-Savart force law.

\subsection{Faraday's law of induction}

\noindent
The interpretation of EM induction within classical electrodynamics makes explicit reference to the EM field, whereas field-free electrodynamics describes the EMF in terms of instantaneous interparticle interactions. We consider a configuration of a primary circuit $ \partial \Omega_2 $ (the boundary of the surface $ \Omega_2 $), which generates an external EM field, and a secondary circuit $ \partial  \Omega_1 $ (the boundary of the surface $ \Omega_1 $), which is influenced by the EM field that is generated by the primary circuit. According to the Lorentz force law, the EMF can be expressed as:
$$   \epsilon  =  \underbrace{ \oint_{ \partial  \Omega_1 }   \textbf{E}  \cdot  d \boldsymbol{\ell}  }_\text{Transformer EMF}    + \underbrace{ \oint_{ \partial \Omega_1 } \textbf{u}  \times  \textbf{B}  \cdot  d \boldsymbol{\ell}  }_\text{Motional EMF}         \hskip \textwidth minus \textwidth  \text{ }  $$
where $ \textbf{u} $ is the velocity of a part of the secondary circuit. Faraday's law encompasses two different phenomena that are linearly superimposed: the transformer EMF caused by an electric force on the secondary circuit, and the motional EMF caused by a magnetic force on the secondary circuit. The total EMF at a given time $t_0 $ can be expressed as:
\begin{flalign}\notag
 \epsilon  [t_0 ] =   \left[ -  \frac{d  \Phi_B  }{ d t  }  \right]_{t = t_0 }  & =  \left[ -  \frac{ d }{ dt }   \int_{  \Omega_1  [t]  }  \textbf{B}     \cdot  d \textbf{A}   \right]_{  t = t_0  }    =    &&  \\\nonumber
& =  \underbrace{  -  \int_{ \Omega_1  [ t_0  ]  }  \left[  \frac{ \partial  \textbf{B} }{  \partial t }  \right]_{ t  = t_0  }   \cdot d  \textbf{A}    }_\text{Transformer EMF}  +  \underbrace{  \left[ -   \frac{d}{dt}  \int_{  \Omega_1  [ t ] }    \textbf{B} [ \textbf{x} ,  t_0 ]  \cdot  d \textbf{A}   \right]_{ t=t_0 }  }_\text{Motional EMF}       && 
\end{flalign}
\textit{Transformer EMF:} According to classical electrodynamics, a changing magnetic field is accompanied by a non-conservative electric field, which induces a current in the secondary circuit. The transformer EMF can be derived using the Maxwell-Faraday equation:
\begin{flalign}\notag
 -  \int_{  \Omega_1 [t_0   ] }  \left[  \frac{ \partial  \textbf{B}  }{ \partial t}  \right]_{  t  = t_0  }   \cdot d  \textbf{A}   &  =     \int_{   \Omega_1  [t_0] }   \left[  \nabla \times   \textbf{E}  \right]_{t = t_0  }  \cdot  d  \textbf{A}    =  \oint_{ \partial   \Omega_1  [ t_0 ] }  \left[ \textbf{E}   \right]_{ t = t_0 }    \cdot   d   \boldsymbol{\ell}          &&
\end{flalign}

\begin{figure}[h]
\centering
\includegraphics[scale=.6]{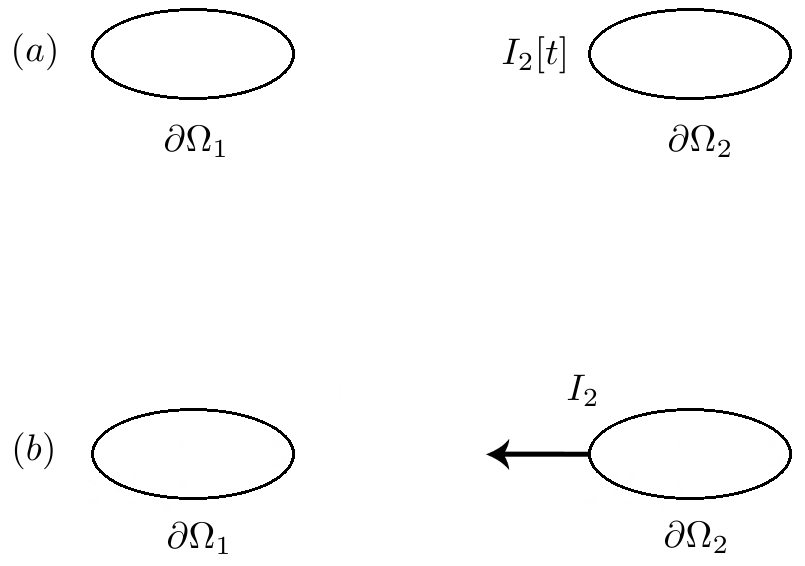}
\caption{An illustration of two kinds of transformer EMFs. (a) The transformer EMF of the fist kind: an intensity variation in the stationary primary circuit $ \partial \Omega_2 $ gives rise to an EMF in the stationary secondary circuit $ \partial \Omega_1 $. (b) The transformer EMF of the second kind: the motion of the current-carrying primary circuit $ \partial \Omega_2 $ results in an EMF in the stationary secondary circuit $ \partial \Omega_1 $.}
\end{figure}

\noindent
We make a distinction between two kinds of transformer EMFs: a transformer EMF caused by an intensity variation in the primary circuit (the transformer EMF of the first kind), and a transformer EMF caused by the motion of a current-carrying primary circuit (the transformer EMF of the second kind). Once again, the transformer EMF in field-free electrodynamics is described in terms of the instantaneous interparticle interactions, and not in terms of the external EM field generated by the primary circuit.

\noindent
\textit{Motional EMF:} A stationary current-carrying primary circuit $ \partial \Omega_2 $ induces a current in a moving secondary circuit $ \partial \Omega_1 $. The time derivative of the magnetic flux through the secondary circuit due to the motion of that circuit is:
$$   \left[  -  \frac{d}{dt }    \int_{ \Omega_1 [t] }  \textbf{B}  [ \textbf{x} , t_0  ] \cdot  \textbf{A}    \right]_{t= t_0 }    =     \oint_{ \partial  \Omega_1 [ t_0 ] }  \left[  \textbf{u}  \times  \textbf{B}  \right]_{ t=t_0 }  \cdot d  \boldsymbol{\ell}       \hskip \textwidth minus \textwidth  \text{ }  $$
The motional EMF will be omitted from the proceeding discussion, because it can be regarded as a consequence of magnetostatics, which has already been covered in Subsection \ref{biot-savart}.

\begin{figure}[h]
\centering
\includegraphics[scale=.6]{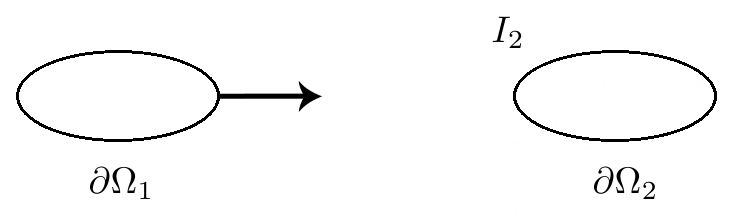}
\caption{An illustration of the motional EMF. The motion of the secondary circuit $ \partial \Omega_1 $ results in an EMF in the same circuit $  \partial \Omega_1 $ in the presence of a stationary current-carrying primary circuit $ \partial \Omega_2 $.}
\label{motional-EMF}
\end{figure}

\subsubsection{Transformer electromotive force in Frejlak's field-free electrodynamics}

\noindent
The electric field at the position $ \textbf{r}_1 $ in the stationary secondary circuit due to a current element $2$ in the primary circuit can be obtained by adding up the $ +$ and $-$ interactions:
\begin{flalign}\notag
    \textbf{E}^{(F,2)}_{12}       & =
 \frac{1}{  e_1  }  \nabla_1    \mathcal{L}^{(F,2)}_{1  2}   -   \frac{1}{  e_1 }  \frac{ d }{    d t}  \left(  \frac{ \partial   \mathcal{L}^{(F,2)}_{1  2 }  }{ \partial  \dot{\textbf{r}}_1 }  \right)   = && \\\nonumber
& =    \frac{1}{  e_1 }   \nabla_1    \mathcal{L}^{(F,2)}_{1  2}    -     \frac{  d }{  d  t}    \left(  k_m    \frac{ I_2  d  \boldsymbol{\ell}_2   +  \hat{\textbf{r}}  (  \hat{\textbf{r}}  \cdot  I_2   d  \boldsymbol{\ell}_2    )   }{ 2 r }  \right)    =    && \\\nonumber
& =  \underbrace{  \frac{1}{  e_1 }   \nabla_1    \mathcal{L}^{(F,2)}_{1  2}  }_{ \substack{ \text{Conservative} \\ \text{vector field}} }  
+ \underbrace{    \frac{ \partial }{ \partial t }  \left(  -  k_m    \frac{ I_2  d  \boldsymbol{\ell}_2   +  \hat{\textbf{r}}  (  \hat{\textbf{r}}  \cdot  I_2   d  \boldsymbol{\ell}_2    )   }{ 2 r }       \right)_{ \textbf{r} }  }_{ \substack{ \text{Transformer EMF} \\ \text{of the first kind}}}   +  \underbrace{       \frac{ \partial }{  \partial t}    \left(  -  k_m    \frac{ I_2  d  \boldsymbol{\ell}_2   +  \hat{\textbf{r}}  (  \hat{\textbf{r}}  \cdot  I_2   d  \boldsymbol{\ell}_2    )   }{ 2 r }  \right)_{ I_2  d  \boldsymbol{\ell}_2  }   }_{ \substack{ \text{Transformer EMF} \\  \text{of the second kind}}}   && 
\end{flalign}
where $ \mathcal{L}^{(F,2)}_{ 1  2} = \mathcal{L}^{(F,2)}_{1  2+}  +  \mathcal{L}^{(F,2)}_{1  2- } $ is Frejlak's field-free Lagrangian of the test charge $e_1$ interacting with current element $2$. The conservative vector field doesn't contribute to the curl of the electric field. The transformer EMF of the first kind is a partial derivative with respect to time while holding $ \textbf{r} $ constant, and the transformer EMF of the second kind is a partial derivative with respect to time while holding $ I_2  d \boldsymbol{\ell}_2 $ constant. Using the identities:
$$   \nabla_1  \times \left(   \dfrac{   \boldsymbol{\alpha}   }{ r }  \right)  =   \dfrac{  \boldsymbol{\alpha}     \times \hat{\textbf{r}} }{  r^2  }       \hskip \textwidth minus \textwidth  \text{ }  $$
$$   \nabla_1   \times  \left(    \dfrac{ \hat{\textbf{r}} ( \hat{\textbf{r}}  \cdot  \boldsymbol{\alpha} )  }{ r }   \right)  =    \dfrac{  \boldsymbol{\alpha}   \times \hat{\textbf{r}}  }{   r^2   }   \hskip \textwidth minus \textwidth  \text{ }  $$
where $ \boldsymbol{\alpha} $ is a constant vector, it follows that the curl of $  \textbf{E}^{(F,2)}_{ 12 } $ is:
$$  \nabla_1  \times \textbf{E}^{(F,2)}_{12}  =  -  \frac{ \partial }{ \partial t}   \left(    \frac{  I_2  d  \boldsymbol{\ell}_2   \times  \hat{\textbf{r}} }{r^2}   \right)      \hskip \textwidth minus \textwidth  \text{ }  $$
We have previously shown that the magnetic field generated by current element $ 2 $ at the position of current element $1$ is given by the Biot-Savart law:
$$    \textbf{B}^{(F,2)}_{12}   =  k_m     \frac{I_2     d  \boldsymbol{\ell}_2   \times \hat{\textbf{r}} }{r^2}       \hskip \textwidth minus \textwidth  \text{ }  $$
from which the Maxwell-Faraday equations follows: $ \dfrac{ \partial  \textbf{B}^{(F,2)}_{12} }{ \partial  t }     =  -  \nabla_1   \times   \textbf{E}^{(F,2)}_{12} $

\subsubsection{Transformer electromotive force in Lorenz's field-free electrodynamics}

\noindent
Let us now turn to Lorenz's field-free electrodynamics. The electric field at the position $ \textbf{r}_1 $ in the stationary secondary circuit due to a current element $2$ in the primary circuit is given by:
\begin{flalign}\notag
    \textbf{E}^{(L,2)}_{12}       & =
 \frac{1}{e_1}  \nabla_1    \mathcal{L}^{(L,2)}_{12}   -  \frac{ d }{ e_1   d   t}  \left(  \frac{ \partial   \mathcal{L}^{(L,2)}_{12}  }{ \partial  \dot{\textbf{r}}_1 }  \right)   = && \\\nonumber
& =    \frac{1}{e_1}   \nabla_1    \mathcal{L}^{(L,2)}_{12}    -     \frac{ d }{  d  t  }    \left(  k_m    \frac{ I_2  d  \boldsymbol{\ell}_2     }{  r }  \right)    =    && \\\nonumber
& =   \underbrace{ \frac{1}{e_1}   \nabla_1    \mathcal{L}^{(L,2)}_{12}  }_{\substack{ \text{Conservative} \\ \text{vector field}}} 
+ \underbrace{    \frac{ \partial }{ \partial t }  \left(  -  k_m    \frac{ I_2  d  \boldsymbol{\ell}_2      }{  r }       \right)_{ \textbf{r} }  }_{\substack{ \text{Transformer EMF} \\ \text{of the first kind}}}   +  \underbrace{       \frac{ \partial  }{  \partial  t  }    \left(  -  k_m    \frac{ I_2  d  \boldsymbol{\ell}_2    }{   r }  \right)_{ I_2  d  \boldsymbol{\ell}_2  }   }_{ \substack{ \text{Transformer EMF} \\  \text{of the second kind} }}  && 
\end{flalign}
The curl of $  \textbf{E}^{(L,2)}_{ 12 } $ is:
$$  \nabla_1  \times \textbf{E}^{(L,2)}_{12}  =  -  \frac{ \partial }{ \partial t}   \left(    \frac{  I_2  d  \boldsymbol{\ell}_2   \times  \hat{\textbf{r}} }{r^2}   \right)        \hskip \textwidth minus \textwidth  \text{ } $$
The magnetic field $  \textbf{B}^{(L,2)}_{1, \partial \Omega_2 }  $ generated by the primary circuit $ \partial \Omega_2 $ at the position of current element $1$ coincides with the Biot-Savart law, so one may also write: $     \textbf{B}^{(L,2)}_{1, \partial \Omega_2 }       = \textbf{B}^{(F,2)}_{1, \partial \Omega_2  }    $. The partial derivative of $ \textbf{B}^{(L,2)}_{1, \partial \Omega_2 } $ with respect to $t$ is:
\begin{flalign}\notag
   \frac{ \partial   \textbf{B}^{(L,2)}_{1 , \partial  \Omega_2 }   }{  \partial t }  & =  
  \oint_{ \partial  \Omega_2  }  \frac{  \partial  \textbf{B}^{( F,2 )}_{12}  }{  \partial  t  }       =   -  \oint_{ \partial  \Omega_2 }   \nabla_1   \times    \textbf{E}^{(L,2)}_{a, 12}     =   -  \nabla_1  \times  \oint_{ \partial  \Omega_2 }         \textbf{E}^{(L,2)}_{a, 12}          &&
\end{flalign}

\newpage
\section{Comparison between Frejlak's and Lorenz's field-free electrodynamics}\label{comparison}

\noindent
Here, we discuss two experiments with which one may distinguish Frejlak's and Lorenz's field-free electrodynamics:

\begin{enumerate}
\item Bound electric field due to steady currents.
\item Weber-type force.
\end{enumerate}

\noindent
\textbf{Experiment 1:} According to Frejlak's field-free electrodynamics, a neutral stationary nonresistive wire that carries a steady current does not generate an electric field: the velocity-dependent electric field is cancelled out by the acceleration-dependent electric field in negotiating the bends, which is a similar situation to what happens in classical electrodynamics.\cite{byrne} Frejlak's field-free Lagrangian of a stationary test charge outside of a neutral wire vanishes: the Coulomb interactions cancel out due to the electrical neutrality of the wire and the velocity-dependent terms vanish. The lack of resistance in the wire ensures that the zeroth and first order electric fields involving the drift velocity $ v_D $ of the conduction electrons are absent from superconductors:\cite{assis-rodriguez}
\begin{enumerate}[-]
\item The zeroth order electric field is due to electrostatic induction.
\item There must be an electric field inside a resistive wire driving the conduction electrons against the resistive forces for a current to flow through a resistive wire. The electric field inside the resistive wire is generated by the surface charges, which also generate an external electric field proportional to $v_D$.
\end{enumerate}
According to Lorenz's field-free electrodynamics, however, a steady current also generates an electric field proportional to $ v_D^2 $, which we refer to as a bound electric field. Note that the existence of a bound electric field is also predicted by Weber's electrodynamics,\cite{assis-weber} which unfortunately has a number of serious shortcomings (Appendix \ref{weber-electrodynamics}). The bound electric field predicted by Lorenz's field-free electrodynamics is highly dependent on the geometry of the conducting wire, as illustrated by the following two configurations:

\begin{enumerate}[-]

\item \textit{Infinite straight current carrying wire:} Consider an infinite straight wire with a uniform and stationary line charge density $+ \lambda $ and a line charge density $   - \lambda $ that moves along the wire with a drift velocity $v_D$. The electric field at a distance $y$ from the wire vanishes: $    \textbf{E}^{(L,2)}     =  0   $.

\item \textit{Magnetic dipole:} Consider a neutral circular current loop $ \mathcal{C} [ \epsilon]$ with a radius $ \epsilon $, positioned at the origin and oriented in the $z$ direction. Once again, let $ v_D $ be the drift velocity of the conduction electrons moving through the current loop. The total electric charge carried by the conduction electrons is $  Q_e $. Lorenz's field-free Lagrangian of a test charge $ q_e $ is:
\begin{flalign}\notag
    \mathcal{L}^{(L,2)}      [  \textbf{r} ]  &  =    -  k_m  q_e   Q_e  v_D^2   \lim_{  \epsilon  \to  0  }  \frac{1}{ 2 \pi \epsilon }  \oint_{ \mathcal{C} [ \epsilon ] }          \frac{ 1 }{ 4  | \textbf{r} -  \textbf{r} '  | }    \left( 1  -   \left(    \frac{ ( \textbf{r}   -  \textbf{r} '  ) }{ |  \textbf{r}  -  \textbf{r} '  | }    \cdot   \left( \hat{\textbf{z}}  \times   \hat{\textbf{r}} '     \right)     \right)^2      \right)   d \textbf{r}'      =  &&  \\\nonumber
& =         -        k_m   \frac{ q_e    Q_e   v_D^2  }{ 8   \pi  r }         \int_0^{2  \pi }       1 -                     \left(      \sin [ \theta ]  \sin [ \varphi  ]      \right)^2      d     \varphi        =  -  k_m   \frac{ q_e    Q_e   v_D^2 }{ 8   r }   \left(   2  -  \sin^2 [ \theta  ]       \right)  &&
\end{flalign}
Hence, the electric field produced by the magnetic dipole is:
$$     \textbf{E}^{(L,2)}   =   \frac{1}{ q_e }   \frac{  \partial  \mathcal{L}  }{  \partial  \textbf{r} }  =
   k_m  \frac{  Q_e  v_D^2  }{8 r^2 }   \left(    \left(  2 -   \sin^2 [ \theta ]    \right)     \hat{\textbf{r}}  +  2  \cos [ \theta ]  \sin [ \theta ]   \hat{\boldsymbol{\theta}}   \right)        \hskip \textwidth minus \textwidth  \text{ }  $$

\end{enumerate}

\noindent
Note that the strength of the bound electric field is negligible in most laboratory conditions, because $ v_D^2$ is typically on the order of $ 10^{-22}$. Once again, the use of superconductors is ideal for the detection of this bound electric field, since the surface charge of the superconducting wire goes to zero (due to the lack of resistance). Hence, the zeroth and first order electric fields are suppressed and only the bound electric field remains. The existence of a bound electric field proportional to $ v_D^2$ was initially reported by Edwards in the 1970s, who used current-carrying superconducting coils to obtain large drift velocities.\cite{edwards1}-\cite{edwards4} Bartlett et al. denied the existence of this effect based on measurements with a rotating solenoid.\cite{bartlett} Finally in the 1990s, Edwards et al. stated that the previously reported results were not fundamental departures from the Maxwell-Lorentz theory of electrodynamics.\cite{lemon} Nevertheless, it would be interesting to repeat the prior experiments in order to determine the possible influence of the geometry of the conducting wire. We mention here some additional experiments and theoretical investigations related to this topic.\cite{bonnet}-\cite{shishkin}

\noindent
\textbf{Experiment 2:} A crucial difference between Frejlak's and Lorenz's field-free electrodynamics is that the latter predicts a Weber-type force that depends on the acceleration of the test body itself. An experiment to distinguish between these competing theories involves an electron accelerated inside a spherical shell with a radius $R$ and an electric charge $ Q_e $ uniformly distributed over its surface. Note that this experimental test was originally proposed as a way to distinguish between classical and Weber's electrodynamics.\cite{assis-weber,assis-mikhailov} According to Frejlak's field-free electrodynamics, the shell does not exert a force on the internal electron. But according to Lorenz's field-free electrodynamics, the electron should behave as if it has an effective inertial mass depending on the surrounding charge. More precisely, the force acting on the electron is:
$$    \textbf{F}^{(L,2)}  =  -   k_e   \frac{ e   Q_e  }{ 4  \pi R^2  }   \int_{  \partial  B [ 0 ,  R  ] }   \frac{ 1}{ 2cr }  \left(      \dot{\boldsymbol{\beta}} -  \hat{\textbf{r}}  ( \hat{\textbf{r}}  \cdot  \dot{\boldsymbol{\beta}}   )   \right)   d  A  =   - k_e   \frac{ e  Q_e }{  3 R c^2   }   \textbf{a}        \hskip \textwidth minus \textwidth  \text{ }  $$
where $  \textbf{a}  $ is the acceleration of the electron. Therefore, the effective inertial mass of the electron is:
$$    m^{(L)}_0   = m_e  + k_e  \frac{e Q_e  }{  3  R c^2  }    \hskip \textwidth minus \textwidth  \text{ }  $$
When the spherical shell is negatively (positively) charged, the inertial mass decreases (increases). An experimental confirmation of a Weber-type force was claimed by Mikhailov.\cite{Mikhailov1}-\cite{mikhailov3} Little et al. reported a null result together with an explanation for Mikhailov's positive result.\cite{little} In particular, they noticed that the measurement of the oscillator's frequency was influenced by the coupling capacitor and observed a null result when they replaced the coupling capacitor with a non-metallic fiber-optic link to a detector located outside the shell.\cite{little} Junginger et al. also performed an experiment employing an optical rather than the electrical coupling and once again reported a null result.\cite{junginger} Recently, the null result has been reproduced in an experiment specifically designed to exclude the possibility that the charges are able to move freely on the surface of the shell, producing image charges and/or induced currents.\cite{tajmar} Another recent reanalysis of Mikhailov's experiment can be found in \cite{tajmar2}.

\newpage
\section{Conclusion}

\noindent
Several shortcomings of classical electrodynamics in a system of point charges can be overcome with IAAD: the interparticle forces are well-defined, there is no diverging energy problem for instantaneous interparticle interactions, there is no infinite regression and there is no immediate threat to causality. IAAD cannot be excluded based on conventional optical experiments, because there is an alternative constructive interpretation of these experiments in terms of IAAD forces that are dependent on the velocity of the measuring device relative to a PRF. We have given three empirically based arguments for reconsidering IAAD:

\begin{enumerate}
\item The dipole anisotropies in the CMB, the galactic red shifts and the muon flux appear to favour the existence of a PRF.
\item Bell's inequality is violated for space-like separated entangled particles.
\item Some experiments call into question the applicability of the standard retardation constraint to all components of the EM field.
\end{enumerate}

\noindent
Special emphasis was placed on Frejlak's and Lorenz's field-free electrodynamics, which are based on the following field-free interaction Lagrangians:
$$   \mathcal{L}^{(F)}_{jk} =   -  k_e  \frac{ e_j  e_k  }{  r }   \frac{1 - \boldsymbol{\beta}_j   \cdot  \boldsymbol{\beta}_k  }{  \sqrt{1-  (  \hat{\textbf{r}}   \times  \boldsymbol{\beta}_j  )  \cdot  ( \hat{\textbf{r}}   \times  \boldsymbol{\beta}_k  )  } }     \hskip \textwidth minus \textwidth  \text{(Frejlak's field-free Lagrangian)}  $$
$$       \mathcal{L}^{(L)}_{jk}      =  -    k_e  \frac{   e_j    e_k  }{ 2 r  }   \left(     \frac{ 1 - \boldsymbol{\beta}_j  \cdot  \boldsymbol{\beta}_k }{  \sqrt{   1-    | \hat{\textbf{r}} \times  \boldsymbol{\beta}_j   |^2     } }   +   \frac{ 1 - \boldsymbol{\beta}_j  \cdot   \boldsymbol{\beta}_k }{  \sqrt{   1-     | \hat{\textbf{r}} \times   \boldsymbol{\beta}_k  |^2     } }        \right)           \hskip \textwidth minus \textwidth  \text{(Lorenz's field-free Lagrangian)}  $$
These Lagrangians incorporate some basic features of classical electrodynamics: the Lorentz transformed Coulomb's law, the Biot-Savart force law and Faraday's law of induction. Lastly, we have proposed two experiments with which one may distinguish Lorenz's and Frejlak's field-free electrodynamics. The first experiment measures the bound electric field of a neutral stationary nonresistive wire that carries a steady current, taking into account the possible influence of the geometry of the conducting wire, and the second experiment involves an isolated charge accelerated inside a charged cage, aimed at detecting a hypothetical Weber-type force. Despite the challenges in measuring the bound electric field and Weber-type force, the preponderance of evidence is consistent with a null result, thus favouring theories that do not exhibit these effects.

\noindent

\newpage
\appendix

\newpage
\section{Magnetostatics in Frejlak's field-free electrodynamics}\label{magnetostatics-frejlak}

\noindent
\begin{tabular}{ | p{7.5cm} | p{7.5cm} | }
\hline
    $ \vphantom{\Bigg(}   \textbf{F}^{(F,2)}_{ 1  \pm, 2  \pm' }  $   &   $   \textbf{F}^{(F,2)}_{12}  =  \textbf{F}^{(F,2)}_{1 + , 2 +} + \textbf{F}^{(F,2)}_{ 1 + , 2 - } + \textbf{F}^{(F,2)}_{ 1 - , 2 + }  + \textbf{F}^{(F,2)}_{ 1 - , 2 - }    $     \\   \hline 
   $ \vphantom{\Bigg(}    k_e  d  e_\pm    d   e_{\pm'}    \dfrac{ \hat{\textbf{r}} }{ r^2  }    $    &      $ 0 $   \\ \hline
   $    -  k_m  d e_{\pm}  d e_{\pm' }   \dfrac{   \hat{\textbf{r}}  }{ 2 r^2 }  (   \dot{\textbf{r}}_{ 1 \pm }  \cdot \dot{\textbf{r}}_{2 \pm' }  )  \vphantom{\Bigg(}  $  & $  - k_m    d  e_+^2  \dfrac{ \hat{\textbf{r}} }{ 2  r^2 }    (  \dot{\textbf{r}}_{1+} - \dot{\textbf{r}}_{1-}  ) \cdot (  \dot{\textbf{r}}_{2+} - \dot{\textbf{r}}_{2-}     ) =  $

$ =  - k_m     I_1 I_2 \dfrac{ \hat{\textbf{r}} }{ 2  r^2}       (  d \boldsymbol{\ell}_1   \cdot  d  \boldsymbol{\ell}_2      )     \vphantom{\Bigg(}  $      \\  \hline
    $   -  k_m   d  e_{ \pm }  d    e_{\pm' }   \dfrac{ 3 \hat{\textbf{r}}}{ 2 r^2 }  ( \hat{\textbf{r}}  \cdot  \dot{\textbf{r}}_{1 \pm}  )( \hat{\textbf{r}}  \cdot  \dot{\textbf{r}}_{2\pm'}   )    \vphantom{\Bigg(}   $  & $     -  k_m     d  e_+^2    \dfrac{ 3  \hat{\textbf{r}} }{ 2  r^2  }   ( \hat{\textbf{r}}  \cdot   ( \dot{\textbf{r}}_{1+} - \dot{\textbf{r}}_{ 1- }  )  ) ( \hat{\textbf{r}} \cdot   (  \dot{\textbf{r}}_{2+}  - \dot{\textbf{r}}_{2-}   ) )   = $

$ =       - k_m   I_1 I_2  \dfrac{ 3 \hat{\textbf{r}}}{ 2  r^2 }  ( \hat{\textbf{r}} \cdot   d  \boldsymbol{\ell}_1  )(  \hat{\textbf{r}} \cdot d \boldsymbol{\ell}_2   )               \vphantom{\Bigg(}      $ \\  \hline
   $    k_m    d  e_{\pm }  d   e_{\pm' }  \dfrac{1}{ 2 r^2 }  \dot{\textbf{r}}_{1 \pm}   (   \hat{\textbf{r}}  \cdot   \dot{\textbf{r}}_{ 2 \pm' }   )  \vphantom{\Bigg(}  $  & $ k_m     d   e_+^2       \dfrac{1}{ 2 r^2 }    ( \dot{\textbf{r}}_{1+} - \dot{\textbf{r}}_{1-}  )(    \hat{\textbf{r}}   \cdot   (  \dot{\textbf{r}}_{2+} -  \dot{\textbf{r}}_{2-}   )    )  = $

$ =   k_m    I_1  I_2    \dfrac{ d  \boldsymbol{\ell}_1   }{ 2 r^2 }    (    \hat{\textbf{r}}   \cdot   d  \boldsymbol{\ell}_2   )  \vphantom{\Bigg(}  $  \\  \hline
  $    k_m   d   e_{ \pm}  d   e_{\pm' }    \dfrac{1}{ 2 r^2 }   \dot{\textbf{r}}_{ 2 \pm' }  (    \hat{\textbf{r}}   \cdot  \dot{\textbf{r}}_{ 1 \pm  }    )  \vphantom{\Bigg(} $   & $   k_m    d   e_+^2    \dfrac{1}{  2  r^2  }     (  \dot{\textbf{r}}_{2+} - \dot{\textbf{r}}_{ 2- }  )(   \hat{\textbf{r}}  \cdot  (  \dot{\textbf{r}}_{1+} -  \dot{\textbf{r}}_{1-}   )    )   = $

$ =  k_m  I_1  I_2     \dfrac{ d  \boldsymbol{\ell}_2   }{ 2 r^2 }     (    \hat{\textbf{r}}  \cdot  d  \boldsymbol{\ell}_1  )      \vphantom{\Bigg(}   $   \\  \hline
 $  -   k_m   d    e_{ \pm}  d  e_{\pm' }    \dfrac{1}{ 2  r }  \ddot{\textbf{r}}_{2 \pm'}      \vphantom{\Bigg(}    $  &     $  0 $    \\    \hline
    $ \vphantom{\Bigg(}    k_m  d   e_{\pm}  d   e_{\pm' }  \dfrac{1}{  2  r^2 } \dot{\textbf{r}}_{ 2 \pm' }  ( \hat{\textbf{r}} \cdot    \dot{\textbf{r}}    )   $ &     $     k_m   d  e_+^2   \dfrac{1}{ 2 r^2 }       (  \dot{\textbf{r}}_{2+} - \dot{\textbf{r}}_{2-} )  ( \hat{\textbf{r}} \cdot  ( \dot{\textbf{r}}_{1+} - \dot{\textbf{r}}_{1-} ) )            = $

$ =   k_m    I_1  I_2  \dfrac{  d  \boldsymbol{\ell}_2   }{2 r^2  }   (  \hat{\textbf{r}} \cdot d  \boldsymbol{\ell}_1   )    \vphantom{\Bigg(} $ \\  \hline
     $  \vphantom{\Bigg(}   k_m  d   e_{\pm} d   e_{\pm' }    \dfrac{ 3  \hat{\textbf{r}} }{ 2   r^2 }       ( \hat{\textbf{r}}  \cdot  \dot{\textbf{r}}_{2 \pm' }  )(  \hat{\textbf{r}}  \cdot   \dot{\textbf{r}}    )      $  &  $      k_m   d  e_+^2  \dfrac{   3   \hat{\textbf{r}}  }{2 r^2 }        (  \hat{\textbf{r}} \cdot  ( \dot{\textbf{r}}_{2+}  -  \dot{\textbf{r}}_{2-}  ) )    ( \hat{\textbf{r}} \cdot  ( \dot{\textbf{r}}_{1+} - \dot{\textbf{r}}_{1-}  ) )           = $

$  =   k_m   I_1  I_2     \dfrac{  3   \hat{\textbf{r}}  }{ 2 r^2 }      (  \hat{\textbf{r}} \cdot   d \boldsymbol{\ell}_2   )  ( \hat{\textbf{r}}  \cdot   d  \boldsymbol{\ell}_1  )                \vphantom{\Bigg(}    $  \\     \hline
   $     -   k_m   d   e_{\pm}   d   e_{\pm' }   \dfrac{ 1 }{ 2   r^2  }    \dot{\textbf{r}}   (  \hat{\textbf{r}} \cdot   \dot{\textbf{r}}_{2 \pm'} )  \vphantom{\Bigg(}  $   &   $       -  k_m  d   e_+^2   \dfrac{ 1 }{ 2 r^2 }      (  \dot{\textbf{r}}_{1+} - \dot{\textbf{r}}_{1-} )  ( \hat{\textbf{r}}  \cdot ( \dot{\textbf{r}}_{2+}  - \dot{\textbf{r}}_{2-} ) )           = $

$ =    -  k_m       I_1  I_2     \dfrac{  d  \boldsymbol{\ell}_1  }{ 2 r^2 }      ( \hat{\textbf{r}}  \cdot  d  \boldsymbol{\ell}_2  )       \vphantom{\Bigg)}     $  \\    \hline
   $    -   k_m    d   e_{\pm}  d   e_{\pm' }  \dfrac{   \hat{\textbf{r}} }{ 2  r^2 }  (    \dot{\textbf{r}}  \cdot  \dot{\textbf{r}}_{2 \pm' }      )  \vphantom{\Bigg(}   $ &    $ -  k_m    d  e_+^2    \dfrac{ \hat{\textbf{r}}}{ 2 r^2 }   ( \dot{\textbf{r}}_{1+} -  \dot{\textbf{r}}_{1-} )  \cdot   (  \dot{\textbf{r}}_{2+} - \dot{\textbf{r}}_{2-} )   =  $

$ =   -   k_m     I_1  I_2  \dfrac{ \hat{\textbf{r}}   }{2 r^2 }  (   d  \boldsymbol{\ell}_1  \cdot   d  \boldsymbol{\ell}_2  )    \vphantom{\Bigg(}  $ \\    \hline
     $ \vphantom{\Bigg(}    -   k_m    d   e_{\pm}  d  e_{\pm' }   \dfrac{ \hat{\textbf{r}}}{2  r }   (  \hat{\textbf{r}}  \cdot  \ddot{\textbf{r}}_{ 2 \pm' }  )  $ &   $   0 $    \\      \hline
    \end{tabular}\\

\noindent
Adding up all the terms in the right column yields the Biot-Savart force law:
\begin{flalign}\notag
   \textbf{F}^{(F,2)}_{12}   & =    -   k_m    I_1  I_2  \frac{  \hat{\textbf{r}} }{  r^2 }         ( d \boldsymbol{\ell}_1  \cdot d \boldsymbol{\ell}_2 )      + k_m    I_1  I_2   \frac{  d \boldsymbol{\ell}_2   }{ r^2}       (     \hat{\textbf{r}}   \cdot   d \boldsymbol{\ell}_1 )     = &&  \\\nonumber
& =    k_m    I_1  I_2   \frac{1}{ r^2 }   d \boldsymbol{\ell}_1  \times ( d  \boldsymbol{\ell}_2  \times  \hat{\textbf{r}}  )     &&
\end{flalign}

\newpage
\section{Magnetostatics in Lorenz's field-free electrodynamics}\label{magnetostatics-lorenz}

\noindent
\begin{tabular}{ | p{7.5cm} | p{7.5cm} | }
\hline
    $ \vphantom{\Bigg(}   \textbf{F}^{(L,2)}_{ 1  \pm, 2  \pm' }  $   &   $   \textbf{F}^{(L,2)}_{12}  =  \textbf{F}^{(L,2)}_{1 + , 2 +} + \textbf{F}^{(L,2)}_{ 1 + , 2 - } + \textbf{F}^{(L,2)}_{ 1 - , 2 + }  + \textbf{F}^{(L,2)}_{ 1 - , 2 - }    $     \\   \hline 
   $ \vphantom{\Bigg(}    k_e  d    e_{ \pm }  d    e_{ \pm' }    \dfrac{ \hat{\textbf{r}} }{ r^2  }    $    &      $ 0 $   \\ \hline
   $    -  k_m  d   e_{\pm}  d   e_{ \pm' }   \dfrac{   \hat{\textbf{r}}  }{  r^2 }  (   \dot{\textbf{r}}_{ 1 \pm }  \cdot \dot{\textbf{r}}_{2 \pm' }  )  \vphantom{\Bigg(}  $  & $  - k_m    d   e_+^2  \dfrac{ \hat{\textbf{r}} }{   r^2 }    (  \dot{\textbf{r}}_{1+} - \dot{\textbf{r}}_{1-}  ) \cdot (  \dot{\textbf{r}}_{2+} - \dot{\textbf{r}}_{2-}     )    =   $

$ =  - k_m     I_1 I_2 \dfrac{ \hat{\textbf{r}} }{   r^2  }       (  d \boldsymbol{\ell}_1   \cdot  d  \boldsymbol{\ell}_2      )     \vphantom{\Bigg(}  $      \\  \hline
    $      k_m   d    e_{ \pm }  d     e_{\pm' }   \dfrac{   \hat{\textbf{r}} }{  4  r^2 }    \left(    \left|  \dot{\textbf{r}}_{ 1 \pm }  \right|^2  +  \left| \dot{\textbf{r}}_{2 \pm '}  \right|^2    \right)           \vphantom{\Bigg(}   $  & $  0     \vphantom{\Bigg(}      $ \\  \hline

    $ -     k_m   d    e_{ \pm }  d    e_{ \pm' }   \dfrac{  3   \hat{\textbf{r}} }{  4  r^2 }    \left(    \left(    \hat{\textbf{r}}  \cdot  \dot{\textbf{r}}_{1 \pm}    \right)^2  +   \left(    \hat{\textbf{r}}  \cdot  \dot{\textbf{r}}_{2 \pm ' }    \right)^2     \right)           \vphantom{\Bigg(}   $  & $  0     \vphantom{\Bigg(}      $ \\  \hline

    $      k_m   d   e_{ \pm }  d    e_{ \pm' }   \dfrac{ 1 }{  2  r^2 }    \left(    \dot{\textbf{r}}_{ 1 \pm } ( \hat{\textbf{r}} \cdot \dot{\textbf{r}}_{1 \pm}  ) 
+   \dot{\textbf{r}}_{ 2 \pm' } ( \hat{\textbf{r}} \cdot \dot{\textbf{r}}_{ 2 \pm' }  )    \right)           \vphantom{\Bigg(}   $  & $  0     \vphantom{\Bigg(}      $ \\  \hline

   $    k_m    d  e_{ \pm }  d  e_{ \pm' }  \dfrac{  1    }{  r^2 }        \dot{\textbf{r}}_{2 \pm '}            ( \hat{\textbf{r}}  \cdot  \dot{\textbf{r}}  )     \vphantom{\Bigg(}  $  & $    k_m  d  e_+^2   \dfrac{  1  }{r^2 }    ( \dot{\textbf{r}}_{2+} - \dot{\textbf{r}}_{2-}  )   (  \hat{\textbf{r}} \cdot (  \dot{\textbf{r}}_{1+} - \dot{\textbf{r}}_{1-} )     )    =  $

$ =   k_m     I_1 I_2  \dfrac{   d  \boldsymbol{\ell}_2  }{   r^2  }       (   \hat{\textbf{r}}   \cdot  d  \boldsymbol{\ell}_1      )     \vphantom{\Bigg(}  $   \\  \hline

$  -   k_m    d   e_{ \pm }  d   e_{ \pm' }  \dfrac{   1 }{ 2  r^2 }     \dot{\textbf{r}}_{1 \pm}     ( \hat{\textbf{r}}  \cdot  \dot{\textbf{r}}    )         \vphantom{\Bigg(}  $  & 
$    k_m  d e_+^2   \dfrac{  1  }{2 r^2 }   (\dot{\textbf{r}}_{1+} - \dot{\textbf{r}}_{1-} )  (  \hat{\textbf{r}}  \cdot ( \dot{\textbf{r}}_{2+}  - \dot{\textbf{r}}_{2-}  )  )     =  $

$ =     k_m     I_1 I_2  \dfrac{   d  \boldsymbol{\ell}_1  }{  2   r^2  }       (   \hat{\textbf{r}}   \cdot  d  \boldsymbol{\ell}_2      )     \vphantom{\Bigg(}  $   \\  \hline

$    k_m    d   e_{ \pm }  d   e_{ \pm' }  \dfrac{ 3   \hat{\textbf{r}}    }{ 2  r^2 }   (  \hat{\textbf{r}}  \cdot  \dot{\textbf{r}}   )    (  \hat{\textbf{r}} \cdot  \dot{\textbf{r}}_{1 \pm }  )         \vphantom{\Bigg(}  $  & 
$    - k_m d   e_+^2  \dfrac{  3   \hat{\textbf{r}}   }{2 r^2  } (  \hat{\textbf{r}} \cdot  (  \dot{\textbf{r}}_{2+} - \dot{\textbf{r}}_{2-}  ) )    (  \hat{\textbf{r}} \cdot  (  \dot{\textbf{r}}_{1+} - \dot{\textbf{r}}_{1-}  ) )  = $

$ =  -   k_m     I_1 I_2  \dfrac{ 3    \hat{\textbf{r}}    }{  2   r^2  }  (   \hat{\textbf{r}}   \cdot  d  \boldsymbol{\ell}_1      )  (   \hat{\textbf{r}}   \cdot  d  \boldsymbol{\ell}_2      )     \vphantom{\Bigg(}  $   \\  \hline

$    k_m    d  e_{ \pm }  d  e_{ \pm' }  \dfrac{1}{  r }  \left(   \dfrac{  \ddot{\textbf{r}}_{1 \pm } }{ 2 }  - \ddot{\textbf{r}}_{2 \pm'}     -    \dfrac{ \hat{\textbf{r}}   (  \hat{\textbf{r}}   \cdot   \ddot{\textbf{r}}_{ 1  \pm  }   )   }{ 2  }    \right)      \vphantom{\Bigg(}  $  & $ 0  $  \\  \hline

$   -   k_m    d  e_{ \pm }  d  e_{ \pm' }   \dfrac{  \hat{\textbf{r}}  }{2 r^2 }  \left(     \dot{\textbf{r}}  \cdot  \dot{\textbf{r}}_{1 \pm }             \right)      \vphantom{\Bigg(}  $  & 
$   k_m  d  e_+^2    \dfrac{ \hat{\textbf{r}} }{ 2 r^2 }  ( \dot{\textbf{r}}_{2+}  -\dot{\textbf{r}}_{2-}  ) \cdot ( \dot{\textbf{r}}_{1+}  -\dot{\textbf{r}}_{1-} )    =   $

$ =   k_m     I_1 I_2  \dfrac{   \hat{\textbf{r}}  }{2   r^2  }       (    d  \boldsymbol{\ell}_1  \cdot  d  \boldsymbol{\ell}_2      )     \vphantom{\Bigg(}  $   \\  \hline

   $  -    k_m    d   e_{ \pm }  d   e_{ \pm' }   \dfrac{1}{ 2 r^2 }         \dot{\textbf{r}}  (    \hat{\textbf{r}}   \cdot   \dot{\textbf{r}}_{1 \pm }   )       \vphantom{\Bigg(}  $  & $  k_m  d  e_+^2   \dfrac{ 1   }{ 2 r^2 }  ( \dot{\textbf{r}}_{2+}  - \dot{\textbf{r}}_{2-}  ) (  \hat{\textbf{r}}  \cdot ( \dot{\textbf{r}}_{1+} - \dot{\textbf{r}}_{1-}  ) )   = $  

$ =   k_m     I_1 I_2  \dfrac{     d  \boldsymbol{\ell}_2     }{ 2  r^2  }          (   \hat{\textbf{r}}  \cdot d \boldsymbol{\ell}_1  )        \vphantom{\Bigg(}  $   \\  \hline

    \end{tabular}\\

\noindent
Adding up all the terms in the right column, yields the following force law:
$$   \textbf{F}^{(L,2)}_{12}    =    -   k_m    I_1  I_2  \frac{  \hat{\textbf{r}} }{  2 r^2 }         ( d  \boldsymbol{\ell}_1  \cdot d \boldsymbol{\ell}_2 )      +
 k_m    I_1  I_2   \frac{  3  d  \boldsymbol{\ell}_2  }{2 r^2}        (    \hat{\textbf{r}} \cdot  d  \boldsymbol{\ell}_1    )   +    
k_m    I_1  I_2   \frac{   d  \boldsymbol{\ell}_1  }{ 2  r^2}        (    \hat{\textbf{r}} \cdot   d  \boldsymbol{\ell}_2   )    
 -    k_m     I_1 I_2  \dfrac{ 3   \hat{\textbf{r}}    }{  2   r^2  }  (   \hat{\textbf{r}}   \cdot  d  \boldsymbol{\ell}_1      )  (   \hat{\textbf{r}}   \cdot  d  \boldsymbol{\ell}_2      )         \hskip \textwidth minus \textwidth  \text{ }  $$

\newpage

\noindent
In order to show its consistency with the Biot-Savart force law when applied to closed circuits, we follow the same proof method that was used by Christodoulides to demonstrate the equivalence between Amp\`{e}re's force law and the Biot-Savart force law.\cite{christodoulides} According to Lorenz's field-free electrodynamics, the force exerted on a current element $1$ by a torus $ V $ is:
$$   \textbf{F}^{(L,2)}_{1  V }    =      k_m   \int_V  -   \frac{  \hat{\textbf{r}} }{  2 r^2 }         (   \textbf{J}_e [ \textbf{r}_1 ]  \cdot    \textbf{J}_e [ \textbf{r}_2  ] )      +        \frac{  3   \textbf{J}_e [ \textbf{r}_2 ]  }{2 r^2}        (     \hat{\textbf{r}}  \cdot    \textbf{J} [ \textbf{r}_1 ] )   +    
 \frac{     \textbf{J}_e [  \textbf{r}_1 ]  }{ 2  r^2}        (    \hat{\textbf{r}}  \cdot  \textbf{J}_e [  \textbf{r}_2  ]  )    
 -         \dfrac{ 3   \hat{\textbf{r}}    }{  2   r^2  }  (   \hat{\textbf{r}}   \cdot    \textbf{J}_e  [ \textbf{r}_1  ]      )  (   \hat{\textbf{r}}   \cdot   \textbf{J}_e   [   \textbf{r}_2  ]     )  d  \textbf{r}_2      \hskip \textwidth minus \textwidth  \text{ }  $$
whereas the Biot-Savart force law yields the following expression:
\begin{flalign}\notag
 \textbf{F}^{(F,2)}_{1   V  }     & =    k_m    \int_{    V }   \frac{  1  }{r^2 }     \textbf{J}_e [ \textbf{r}_1 ]  \times (   \textbf{J}_e [ \textbf{r}_2  ]   \times   \hat{\textbf{r}}  )   d \textbf{r}_2       =   && \\\nonumber
&  =   k_m    \int_{  V  }   \frac{  1  }{r^2 }     (     \textbf{J}_e [ \textbf{r}_2 ]  (   \hat{\textbf{r}} \cdot  \textbf{J}_e [ \textbf{r}_1 ]    )   -    \hat{\textbf{r}}   (  \textbf{J}_e  [ \textbf{r}_1  ] \cdot    \textbf{J}_e  [ \textbf{r}_2 ]  )    )   d   \textbf{r}_2     &&  
  \end{flalign}
The torus $V$ is bounded by a closed surface $\partial V$. It is assumed that $\textbf{J}_e  \cdot  \hat{\textbf{n}} = 0$, where $ \hat{\textbf{n}}$ is the unit normal vector of the surface $\partial V$. We wish to show that the difference between $   \textbf{F}^{(L,2)}_{1  V }  $ and $  \textbf{F}^{(F,2)}_{1  V } $ vanishes:
$$          \int_V    \frac{  \hat{\textbf{r}} }{  2 r^2 }         (   \textbf{J}_e [ \textbf{r}_1 ]  \cdot    \textbf{J}_e [ \textbf{r}_2  ] )      +        \frac{    \textbf{J}_e [ \textbf{r}_2 ]  }{2 r^2}        (    \hat{\textbf{r}}  \cdot  \textbf{J} [ \textbf{r}_1 ]   )   +     \frac{     \textbf{J}_e [  \textbf{r}_1 ]  }{ 2  r^2}        (     \hat{\textbf{r}} \cdot \textbf{J}_e [  \textbf{r}_2  ]   )    
 -         \dfrac{ 3   \hat{\textbf{r}}    }{  2   r^2  }  (   \hat{\textbf{r}}   \cdot    \textbf{J}_e  [ \textbf{r}_1  ]      )  (   \hat{\textbf{r}}   \cdot   \textbf{J}_e   [   \textbf{r}_2  ]     )  d  \textbf{r}_2    = 0    \hskip \textwidth minus \textwidth  \text{ }  $$

\begin{figure}[h]
\centering
\includegraphics[scale=.6]{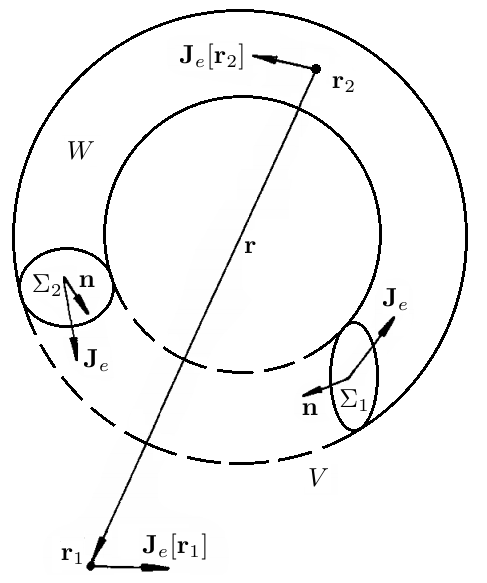}
\caption{The geometry of the electric current distribution $\textbf{J}_e$.}
\label{magnetostatics}
\end{figure}

\noindent
Using $ \nabla_2 \times \textbf{r}    = 0 $ and $ \nabla_2 \cdot  \textbf{J}_e  [\textbf{r}_2 ] = 0$, we find:
$$  \nabla_2  \cdot  \left(  \frac{ \textbf{J}_e [ \textbf{r}_2 ] }{ r^2 }   (       \hat{\textbf{r}} \cdot  \textbf{J}_e  [ \textbf{r}_1 ]  )  \right)  = 
 \frac{ \textbf{J}_e [  \textbf{r}_1 ] \cdot \textbf{J}_e [ \textbf{r}_2 ] }{ r^3  }   - \frac{3 }{r^3 }  (  \hat{\textbf{r}}   \cdot   \textbf{J}_e [  \textbf{r}_1 ]  )(   \hat{\textbf{r}}    \cdot \textbf{J}_e [ \textbf{r}_2 ] )  
 \hskip \textwidth minus \textwidth  \text{ }  $$
So we wish to prove:
$$    \int_V   \textbf{N}   +  \textbf{r}    \left(  \nabla_2 \cdot  \textbf{N}  \right)    d \textbf{r}_2   +     \int_V  \frac{ \textbf{J}_e  [ \textbf{r}_1  ] }{ r^2 }    (       \hat{\textbf{r}}  \cdot   \textbf{J}_e  [ \textbf{r}_ 2 ]      )  d \textbf{r}_2        =  0    \hskip \textwidth minus \textwidth  \text{ }  $$
where: $ \textbf{N} =      \textbf{J}_e [ \textbf{r}_2 ]    (     \hat{\textbf{r}}  \cdot   \textbf{J}_e  [ \textbf{r}_1 ]   )  /   r^2  $. In fact, it will be shown that both integrals vanish:
$$    \int_V   \textbf{N}   +  \textbf{r}    \left(  \nabla_2 \cdot  \textbf{N}  \right)    d \textbf{r}_2      =  0    \hskip \textwidth minus \textwidth  \text{ }  $$
$$        \int_V  \frac{       \hat{\textbf{r}}   \cdot   \textbf{J}_e  [ \textbf{r}_ 2 ]    }{ r^2 }     d \textbf{r}_2        =  0    \hskip \textwidth minus \textwidth  \text{ }  $$
The first integral over the volume $W \subseteq  V$ can be written as an integral over a surface $\partial W$ enclosing $W$:
\begin{flalign}\notag
     \int_{W}  \textbf{N}   +  \textbf{r}  ( \nabla_2 \cdot  \textbf{N} )  d  \textbf{r}_2   &  =  \int_{W}  ( \textbf{N} \cdot \nabla_2 ) \textbf{r}  +  \textbf{r}   ( \nabla_2   \cdot \textbf{N} )   d \textbf{r}_2    =  &&  \\\nonumber
&  =   \int_{ \partial W}  \textbf{r} (   \hat{\textbf{n}}  \cdot   \textbf{N}   )    d  \textbf{r}_2     = 
   &&   \\\nonumber
&  =  \int_{ \partial W}  \frac{   \hat{\textbf{r}}  }{    r  } (    \hat{\textbf{n}}   \cdot   \textbf{J}_e [ \textbf{r}_2 ] )  (    \hat{\textbf{r}}  \cdot   \textbf{J}_e [ \textbf{r}_1 ]   )  d  \textbf{r}_2     &&
  \end{flalign}
Once again, $ \hat{\textbf{n}}$ is the outward unit normal vector on the surface at the position $\textbf{r}_2$. For any $ \textbf{r}_2 \in \partial V $, we have: $    \hat{\textbf{n}}  \cdot   \textbf{J}_e [ \textbf{r}_2 ]  = 0  $. Hence, this integral can be written as:
$$   \int_{ \partial W}  \frac{   \hat{\textbf{r}}   }{  r } (   \hat{\textbf{n}}  \cdot   \textbf{J}_e [ \textbf{r}_2 ]   )  (   \hat{\textbf{r}}    \cdot \textbf{J}_e [ \textbf{r}_1 ]  )  d  \textbf{r}_2   =       \int_{ \Sigma_1  }   \frac{  \hat{\textbf{r}}  }{ r  } (   \hat{\textbf{n}}  \cdot \textbf{J}_e [ \textbf{r}_2 ]   )  (    \hat{\textbf{r}}   \cdot   \textbf{J}_e [ \textbf{r}_1 ]    )  d  \textbf{r}_2    +  \int_{ \Sigma_2  }   \frac{   \hat{\textbf{r}}  }{  r  } (  \hat{\textbf{n}}  \cdot   \textbf{J}_e [ \textbf{r}_2 ]  )  (    \hat{\textbf{r}}  \cdot \textbf{J}_e [ \textbf{r}_1 ]   )  d  \textbf{r}_2     \hskip \textwidth minus \textwidth  \text{ }  $$
If $\Sigma_1 =  \Sigma_2$, completing the current loop, then the integral vanishes, since the unit normal vectors have opposite directions. The same method can be applied to the second integral:
\begin{flalign}\notag
  \int_W   \frac{     \hat{\textbf{r}}   \cdot \textbf{J}_e [ \textbf{r}_2  ]        }{ r^2  }     d \textbf{r}_2   & =   
   \int_W     \nabla_2  \cdot  \left(   \frac{  \textbf{J}_e  [  \textbf{r}_2  ]  }{ r }   \right)        d \textbf{r}_2        = &&  \\\nonumber
& =         \int_{ \partial W }    \frac{   \hat{\textbf{n}}  \cdot \textbf{J}_e  [ \textbf{r}_2 ]     }{ r }  d  \textbf{r}_2  =   && \\\nonumber
& =    \int_{ \Sigma_1  }    \frac{     \hat{\textbf{n}} \cdot  \textbf{J}_e  [ \textbf{r}_2 ]     }{ r }  d \textbf{r}_2    + 
 \int_{ \Sigma_2  }   \frac{     \hat{\textbf{n}}   \cdot    \textbf{J}_e  [ \textbf{r}_2 ]     }{ r }   d \textbf{r}_2    &&
\end{flalign}
If $ \Sigma_1 = \Sigma_2  $, then the integral vanishes and we conclude that $ \textbf{F}^{(L,2)}_{1V} = \textbf{F}^{(F,2)}_{1V} $.

\newpage
\section{Generalized field-free interaction Lagrangian}\label{generalized-lagrangian}

\noindent
Suppose that the general expression for the field-free interaction Lagrangian including terms up to the second order in $1/c$ is given by:
$$      \mathcal{L}_{jk}^{(2)} =   - k_e    \frac{ e_j   e_k }{r}   \left(  1 +  \alpha_1  \boldsymbol{\beta}_j  \cdot  \boldsymbol{\beta}_k    +  \alpha_2   ( \beta_j^2  +  \beta_k^2  )  +  \alpha_3   (  \hat{\textbf{r}}  \cdot  \boldsymbol{\beta}_j  )   (  \hat{\textbf{r}}  \cdot  \boldsymbol{\beta}_k  )  + \alpha_4  (  ( \hat{\textbf{r}} \cdot  \boldsymbol{\beta}_j )^2    +   (  \hat{\textbf{r}}  \cdot  \boldsymbol{\beta}_k   )^2 )    \right)     \hskip \textwidth minus \textwidth  \text{ }  $$
which is symmetric under exchange of the coordinates of the particles. We determine the conditions under which this generalized Lagrangian shows consistency with the following two force laws:

\begin{enumerate}[-]
\item \textit{Lorentz transformed Coulomb's law:} The low-velocity limit of this force law is:
$$    \textbf{F}^{(2)}_{jk}   =   k_e   e_j   e_k   \frac{ \hat{\textbf{r}} }{r^2 }    \left( 1 -   \frac{ \beta^2 }{2}   - \frac{3}{2} ( \hat{\textbf{r}}  \cdot  \boldsymbol{\beta} )^2  \right)  + k_e   e_j  e_k  \frac{1}{r^2 }  \boldsymbol{\beta} ( \hat{\textbf{r}} \cdot  \boldsymbol{\beta} )    \hskip \textwidth minus \textwidth  \text{ }  $$
The force between two uniformly moving point charges with the same velocity corresponding to the generalized Lagrangian is:
$$     \textbf{F}^{(2)}_{jk}   =  \frac{ \partial  \mathcal{L}_{jk}^{(2)} }{ \partial  \textbf{r}_j  }  =   k_e   e_j   e_k     \frac{ \hat{\textbf{r}}  }{r^2}   \left(  1  + ( \alpha_1 + 2 \alpha_2  )  \beta^2        +  
3 ( \alpha_3   +  2  \alpha_4 ) (  \hat{\textbf{r}}  \cdot   \boldsymbol{\beta}  )^2       \right)    -    k_e  e_j  e_k   \frac{1}{ r^2}   2 ( \alpha_3  + 2  \alpha_4 )   \boldsymbol{\beta} ( \hat{\textbf{r}}  \cdot \boldsymbol{\beta}   )              \hskip \textwidth minus \textwidth  \text{ }  $$
which shows consistency with the Lorentz transformed Coulomb's law if the following conditions hold:
$$    \alpha_1 +  2  \alpha_2   =   - 1/2             \hskip \textwidth minus \textwidth  \text{(Condition 1.1)}  $$
$$    \alpha_3  +  2  \alpha_4   =   - 1/2             \hskip \textwidth minus \textwidth  \text{(Condition 1.2)}  $$

\item \textit{Biot-Savart force law:} The force between two current elements corresponding to the generalized Lagrangian can be obtained by adding up the $++$, $+-$, $-+$ and $--$ interactions:
\begin{flalign}\notag
   \textbf{F}^{(2)}_{12}  & =  k_m I_1  I_2  \frac{ \hat{\textbf{r}} }{r^2} ( \alpha_1 +  \alpha_3 - 2  \alpha_4 ) (  d \boldsymbol{\ell}_1  \cdot d  \boldsymbol{\ell}_2 )    -  k_m  I_1  I_2  \frac{ d \boldsymbol{\ell}_2  }{ r^2 }  ( \alpha_1 +  \alpha_3  + 2  \alpha_4  )  ( \hat{\textbf{r}}  \cdot d  \boldsymbol{\ell}_1 )     +      &&    \\\nonumber
& +  k_m  I_1  I_2    \frac{ d \boldsymbol{\ell}_1 }{r^2} 2 \alpha_2   ( \hat{\textbf{r}} \cdot  \boldsymbol{\ell}_2 )   +    k_m I_1  I_2   \frac{   \hat{\textbf{r}} }{ r^2 }   6   \alpha_4   ( \hat{\textbf{r}}  \cdot  d \boldsymbol{\ell}_1  )( \hat{\textbf{r}} \cdot  d  \boldsymbol{\ell}_2  )  &&  
\end{flalign}
which shows consistency with the Biot-Savart force law if the following condition holds:
$$    \alpha_1 +    \alpha_3   =     -1             \hskip \textwidth minus \textwidth  \text{(Condition 2)}  $$
\end{enumerate}

\noindent
The coefficients corresponding to $ \mathcal{L}^{(F,2)}_{jk} $ and $ \mathcal{L}^{(L,2)}_{jk}  $ are shown in the following table:

\noindent
\begin{tabular}{ |  p{2.5cm}  | p{2.5cm} | p{2.5cm} |   p{2.5cm}  |  p{2.5cm} | }
    \hline 
   $ \vphantom{\Bigg(} $     &   $  \alpha_1 $    &   $ \alpha_2 $ &  $ \alpha_3 $   &  $  \alpha_4 $ \\ \hline
    $  \mathcal{L}^{(F,2)}_{jk} $   &  $ \vphantom{\Bigg(}  -  1/ 2   $   &   $    0  $  & $- 1/2$  &  $0$  \\ \hline
      $  \mathcal{L}^{(L,2)}_{jk} $   &  $ \vphantom{\Bigg(}   -1  $   &   $   1/4  $  &   $0$  &  $  - 1/4  $ \\ \hline
    \end{tabular}

\newpage
\section{Weber's electrodynamics}\label{weber-electrodynamics}

\noindent
Weber's electrodynamics is a relational theory of electrodynamics based on the following field-free interaction Lagrangian:\cite{assis-weber}
$$   \mathcal{L}^{(W)}_{jk} =    - k_e  \frac{ q_{e1} q_{e2}  }{  r}   \left(  1 + \frac{ \dot{r}^2 }{2 c^2 } \right)    \hskip \textwidth minus \textwidth  \text{(Weber's field-free Lagrangian)}  $$
This Lagrangian is relational in that it depends on relational quantities such as the distance $r$ between the interacting particles and the relative radial velocity $\dot{r}$ between them. If there is no relative motion between two interacting electric charges, we recover Coulomb's force law. And the force between two neutral current elements is given by Amp\`{e}re's force law:
$$  \textbf{F}^{(W)}_{12}  =   k_m  \frac{ \hat{\textbf{r}} }{ r^2 }  I_1  I_2    ( 3  ( \hat{\textbf{r}} \cdot  d \boldsymbol{\ell}_1 ) (   \hat{\textbf{r}}   \cdot  d \boldsymbol{\ell}_2  )  -  2 ( d \boldsymbol{\ell}_1  \cdot  d \boldsymbol{\ell}_2 )     )    \hskip \textwidth minus \textwidth  \text{ }  $$
which is equivalent to the Biot-Savart force law when applied to closed circuits.\cite{christodoulides} However, Weber's electrodynamics has a number of undesirable features:

\begin{enumerate}[-]
\item Weber's electrodynamics has never been extended to the high-velocity regime and its purely relational character suggests that it is incompatible with the Lorentz transformed Coulomb's law. Indeed, the low-velocity limit of the Lorentz transformed Coulomb's law yields an expression that depends not only on relational quantities, but also on the absolute motion of the interacting particles. The coefficients of the generalized field-free interaction Lagrangian (as presented in Appendix \ref{generalized-lagrangian}) corresponding to Weber's field-free Lagrangian are shown in the following table:

\noindent
\begin{tabular}{ |  p{2.5cm}  | p{2.5cm} | p{2.5cm} |   p{2.5cm}  |  p{2.5cm} | }
    \hline 
   $ \vphantom{\Bigg(} $     &   $  \alpha_1 $    &   $ \alpha_2 $ &  $ \alpha_3 $   &  $  \alpha_4 $ \\ \hline
    $  \mathcal{L}^{(W)}_{jk} $   &  $ \vphantom{\Bigg(}   $0$   $   &   $    0  $  & $-1$  &  $  1/2$  \\ \hline
    \end{tabular}

\noindent
which satisfies Condition 2, but contradicts Condition 1.1 and Condition 1.2.

\item The Bucherer experiment aimed to measure the dependence of the transverse mass of an electron on its velocity. Bush and O'Rahilly were the first to apply Weber's electrodynamics to the Bucherer experiment.\cite{bush,orahilly} This calculation was later rediscovered by Wesley and Assis.\cite{wesley3,assis-bucherer} Weber's electrodynamics yields the correct radius of the electron's circular path up to the second order without having to make the usual assumption of a transverse mass increase. However, plenty of experiments have demonstrated the dependence of mass on velocity in a direct way (e.g. the Bertozzi experiment \cite{bertozzi}). Furthermore, relativistic kinematics is applicable to high-velocity particle collisions. Hence, one may turn the argument on its head: once the mass variation is taken into account, the Bucherer experiment is no longer consistent with Weber's electrodynamics.

\item Lastly, Weber's electrodynamics (just like Lorenz's field-free electrodynamics) exhibits a bound electric field and a Weber-type force. Despite the challenges in measuring these effects, the preponderance of evidence appears to be consistent with a null result (Section \ref{comparison}).
\end{enumerate}

\end{document}